\pdfoutput=1
\documentclass[jgrga]{agutex}

\usepackage{lineno}
\usepackage{epsfig}
\usepackage{graphicx}

\authorrunninghead{DORELLI ET AL.}

\titlerunninghead{Hall MHD Ganymede Simulations}

\authoraddr{Corresponding author: J. C. Dorelli,
Geospace Physics Laboratory, NASA Goddard Space Flight Center, 
Mail Code 673, Greenbelt, MD 20851, USA.
(john.dorelli@nasa.gov)}

\begin{document}

\title{The role of the Hall effect in the global structure and dynamics of planetary magnetospheres:  Ganymede as a case study}

 \authors{J. C. Dorelli,\altaffilmark{1}
 Alex Glocer, \altaffilmark{1}
 Glyn Collinson, \altaffilmark{1}
 G\'{a}bor T\'{o}th \altaffilmark{2}}

\altaffiltext{1}{Geospace Physics Laboratory,
NASA-GSFC, Greenbelt, Maryland, USA.}

\altaffiltext{2}{Center for Space Environment Modeling,
University of Michigan, Ann Arbor, Michigan, USA.}

\begin{abstract}
We present high resolution Hall MHD simulations of Ganymede's magnetosphere demonstrating that Hall electric fields in ion-scale magnetic reconnection layers have significant global effects not captured in resistive MHD simulations.
Consistent with local kinetic simulations of magnetic reconnection, our global simulations show the development of intense field-aligned currents along the magnetic separatrices.
These currents extend all the way down to the moon's surface, where they may contribute to Ganymede's aurora.
Within the magnetopause and magnetotail current sheets, Hall currents in the reconnection plane accelerate ions to the local Alfv\'{e}n speed in the out-of-plane direction, producing a global system of ion drift belts that circulates Jovian magnetospheric plasma throughout Ganymede's magnetosphere.
We discuss some observable consequences of these Hall-induced currents and ion drifts:  the appearance of a sub-Jovian "double magnetopause" structure, an Alfv\'{e}nic ion jet extending across the upstream magnetopause and an asymmetric pattern of magnetopause Kelvin-Helmholtz waves.
\end{abstract}

\begin{article}

\section{Introduction \label{SecIntroduction}}

The rapid reconnection-driven response of magnetospheric convection to changes in the orientation of the external magnetic field is one of the great puzzles of magnetospheric physics.
The nearly collisionless magnetospheric plasma should effectively shield it from changes in the external field.
In the resistive mangetohydrodynamics (MHD) limit, magnetospheric reconnection should occur in vanishingly thin extended current sheets on time scales much too slow \citep{parkerd,sweeta} to explain magnetic storms and substorms.
Either there is some anomalous plasma resistivity (e.g., produced by turbulence) enhancing the Sweet-Parker reconnection rate, or higher-order terms in the Generalized Ohm's Law (e.g., the Hall effect, electron pressure anisotropy, or electron inertia) are somehow preventing the formation of extended electron-scale current sheets.

The important role of Hall electric fields in allowing fast reconnection to occur in the limit of very small plasma resistivity was only beginning to be appreciated in the late 90's \citep{biskampc,biskampa,mac,shayb,shaya}, culminating in the ``GEM (Geospace Environment Modeling) Reconnection Challenge" results summarized in \citet{birnc}.
While there is still debate about the specific physical mechanisms involved -- e.g., Are the dispersive properties of whistler and kinetic Alfv\'{e}n waves playing the dominant role \citep{mandta,rogersa}?  Are finite Larmor radius effects playing a crucial part \citep{hessea,hesseb}? -- there is general agreement that simply including the Hall term in the Generalized Ohm's Law is a huge improvement over resistive MHD, producing reconnection rates that are comparable to those observed in full particle-in-cell (PIC) simulations.

It is less clear how the GEM Reconnection Challenge results ``scale up" to very large systems like planetary magnetospheres or stellar coronal active regions, whose characteristic scales $L$ are many orders of magnitude larger than the Larmor radii or inertial scale lengths over which collisionless reconnection physics dominates.
While there was some early evidence suggesting that Hall reconnection produced Petschek-like \citep{petscheka} configurations that rendered the reconnection rate insensitive to the system size \citep{shaya}, subsequent Hall MHD simulations of magnetic island coalescence -- in which reconnection is driven by the ideal MHD instability of a system of magnetic structures with characteristic scale $\lambda$ (the island wavelength) -- suggested that the reconnection rate (both the instantaneous rate and the coalescence time) should scale like $(d_i/\lambda)^{1/2}$, where $d_i$ is the ion inertial length \citep{dorellic,dorellid,knolla}.
Recent electromagnetic PIC simulations \citep{karimabadic} of large scale island coalescence are consistent with the earlier Hall MHD results in some ways (predicting a coalescence time that scales like $(d_i/\lambda)^{1/2}$), but they nevertheless support the idea that the maximum instantaneous reconnection rate is insensitive to the island size (at least over the limited parameter range so far accessible by PIC simulations).

Moving from two to three dimensions produces a much richer reconnection landscape.
Breaking the two-dimensional symmetry permits plasma instabilities (e.g., oblique tearing modes \citep{galeeva,daughtond}) that would have been suppressed in two-dimensions.
Recently, \citet{daughtond} have argued that collisionless magnetic reconnection in a realistic three-dimensional magnetosphere should produce extended electron-scale current sheets that in turn become unstable to secondary flux rope formation, ultimately resulting in a turbulent reconnection layer.

Unfortunately, high performance computing capabilities are not yet advanced enough to permit even Hall MHD simulations of an Earth-sized magnetosphere; the same dispersive waves that may play a role in producing fast reconnection also make the Courant-Lewy-Friedrichs time step prohibitively small for explicit codes.
Thus, it is not surprising that most Earth-scale global magnetosphere simulations to date still use resistive MHD, either relying on numerical resistivity or {\it ad hoc} current-dependent ``anomalous" resistivity to produce fast reconnection.
While it has been known since the GEM Reconnection Challenge that one can easily achieve reconnection rates comparable with observations and kinetic simulations by using current-dependent resistivity models with appropriately dialed free parameters (e.g., \citet{ottoa}), it is by no means clear that simply getting the right local reconnection rate is the only way that kinetic scale physics influences global magnetospheric structure and dynamics.
In this work, we address this basic question:  Does the local structure of collisionless magnetic reconnection (specifically, the Hall current structure within the ion inertial region) influence the global structure and dynamics (viz., convection and field-aligned current patterns) of planetary magnetospheres?

While Earth may still be out of reach of present day Hall MHD codes -- and even Mercury pushes the high performance computational envelope -- Jupiter's third Galilean moon, Ganymede, provides us with an ideal opportunity to begin addressing these questions. 
The distance between the surface of the moon and the upstream magnetopause is about a factor of $10$ larger than the oxygen ion skin depth in the surrounding Jovian magnetosphere (e.g., \citet{kivelsona}), providing us with a $d_i/L << 1$ magnetosphere that is still computationally tractable.
Ganymede is also unique in that the Jovian magnetospheric magnetic field is strongly anti-aligned with Ganymede's dipole, so that magnetopause reconnection is strongly driven by nearly steady upstream conditions, a situation analogous to the conditions that drive enhanced magnetospheric convection and magnetic storms in Earth's magnetosphere.
Finally, we have several Galileo flybys under essentially steady magnetospheric conditions, allowing for a relatively straightforward observational test of our simulations.

Recent high resolution resistive MHD simulations of Ganymede's magnetosphere by \citet{jiaa, jiab} have produced a global picture of Ganymede's magnetic field that agrees well with Galileo magnetometer observations for the six Ganymede flybys (e.g., reproducing such basic features as the location and shape of the magnetopause).
While this may seem at first glance to support the idea that ion-scale physics does not have a significant impact on Ganymede's global magnetic field structure, it is important to note that only two of the flybys (G8 and G28) crossed the upstream magnetopause at locations close enough to the reconnection site to test predictions about global convection and field-aligned current patterns.
The G7 and G29 flybys did not sample the central tail current sheet (where, as we will see below, most of the reconnection-driven convection is confined), and the G1 flyby was too close to the moon's surface to sample the tail current sheet.
Thus, simply comparing a simulation to a few Galileo flybys is not sufficient to address the impact of ion-scale physics on global magnetospheric structure; MHD may get the basic size and shape of the magnetopause right and still (as we will demonstrate below) fail to correctly capture the global convection and field-aligned current patterns.

In what follows, we describe our use of global Hall MHD simulations of Ganymede's magnetosphere to demonstrate that the local ion-scale structures observed in the GEM Reconnection Challenge (and all subsequent kinetic reconnection simulations) have significant impacts on Ganymede's convection and field-aligned current patterns.
Specifically, the Hall ``out-of-plane" magnetic field quadrupole pattern that has become one of the defining characteristics of collisionless magnetic reconnection generates a new system of field-aligned currents that map directly from the reconnection sites all the way down to the moon's surface, modifying the region-1 type current system that supports the Alfv\'{e}n wing structure close to the moon.
Further, the same Hall current system produces a new ${\bf J} \times {\bf B}$ force that accelerates ions to their local Alfv\'{e}n speed out of the reconnection plane, introducing large asymmetries to the classic \citet{dungeya} convection pattern.

In previous pioneering work, \citet{wingleea,wingleeb} emphasized some of the effects described in this paper (in particular, the impact of Hall physics on the field-aligned current systems and the important role of non-MHD perpendicular ion drifts).
However, the picture that emerged from this previous work was incomplete in important ways.
The particle simulations of \citet{wingleea} were two-dimensional and did not properly capture the impact of non-MHD drifts on the three-dimensional convection pattern (as we will demonstrate below).
The \citet{wingleeb} simulations, while three-dimensional, used cell sizes a factor of $\sim 8$ larger than the ion inertial length, implying that ion scale effects were likely swamped by numerical dissipation.
As we will see below, it is crucial that one resolve the ion inertial length with at least $5$ computational cells per ion inertial length in three dimensions to see significant differences between the resistive and Hall MHD global convection patterns.

Previous multi-fluid simulations of Ganymede \citep{patya,patyb,patyc} that included the Hall effect have achieved resolution comparable to that of the simulations described here; however, this previous work focused on the effects of ionospheric outlow (separately modeling the hydrogen and oxygen populations) and did not discuss the impact of Hall-mediated reconnection on the global pattern of field-aligned currents and convection.
Further, it is not clear from these simulations how the enhanced ionospheric outflow impacted the ability to resolve the ion inertial length:  $d_i \sim n_{ion}^{-1/2}$, implying that more oxygen flowing out into the magnetosphere makes it more difficult to achieve $>5$ computational cells per $d_i$.
In contrast, the present work does not explore the role of ionospheric outflow, focusing instead on the global effects of ion-scale reconnection (which are much more difficult to resolve in the presence of significant ion outflow).
We plan to explore the effect of ionospheric outflow in future work using the multi-fluid models developed by \citet{glocera}.

\section{Simulation setup \label{SecSimulationParameters}}

\begin{figure*}[th]
\centering
\includegraphics[width=6in,height=2.5in]{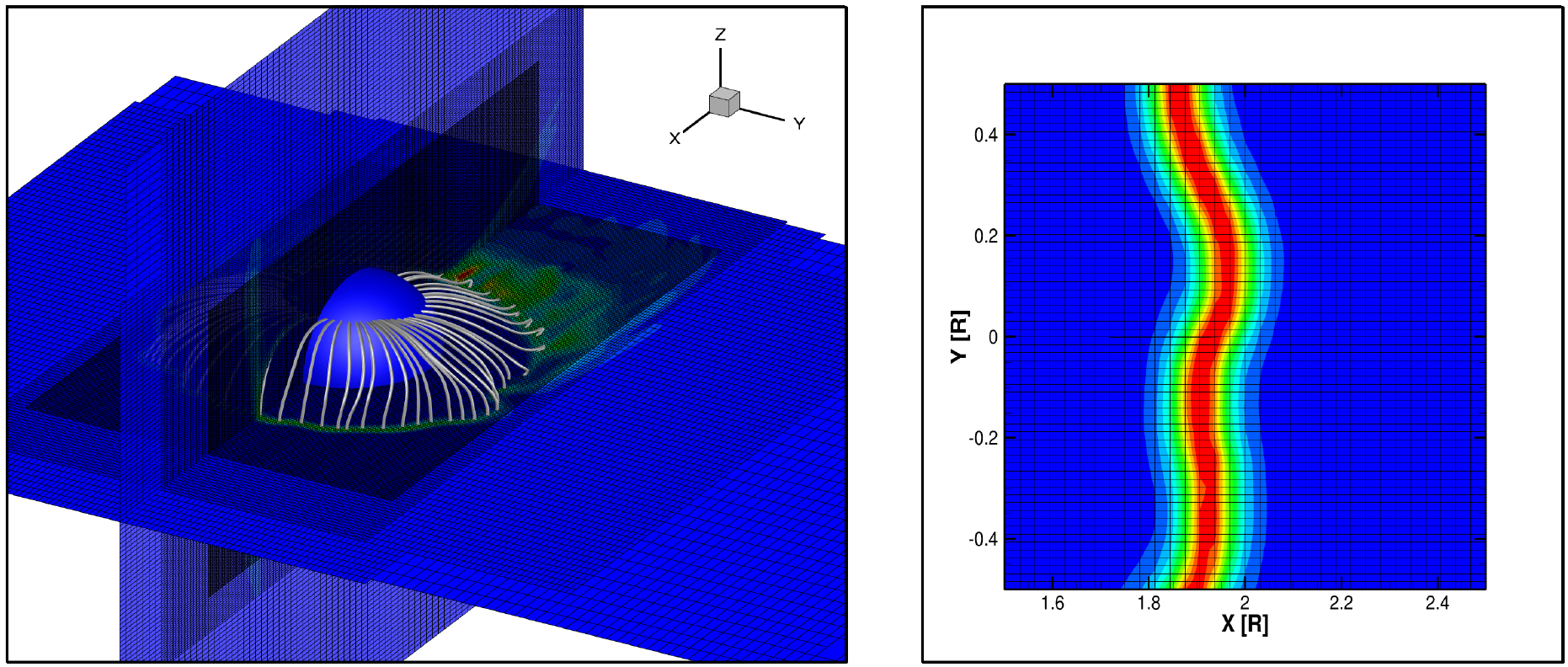}
\caption{\label{FigGrid} Hall effects appear when the dissipation scale is sufficiently smaller than $d_i$, so that the ion and electron bulk velocities become decoupled in thin $d_i$ scale current sheets.
To achieve this scale separation, while still recovering the large scale Alfv\'{e}n wing structure at large distances from the moon, we used a nested grid in which the highest resolution of $1/32 R_G$ was concentrated near the moon (left panel).
This allowed us to achieve a resolution of $5-10$ computational cells within the $d_i$ scale current sheet (right panel).
Thus, since the numerical scheme concentrates numerical resistivity at the grid scale, this strategy achieves the required separation between $d_i$ and the dissipation scale.}
\end{figure*}

We performed our simulations using the BATS-R-US global Hall magnetohydrodynamics code developed at the University of Michigan \citep{powella,totha}.
The Jovian magnetospheric plasma and field parameters were the same as those used by \citet{jiaa} for their simulation of the Galileo G8 flyby.
The Jovian magnetospheric plasma and field parameters were the same as those used by \citet{jiaa} for their simulation of the Galileo G8 flyby.
Ganymede's dipole moment was set to  $M_X = -716.8 \, nT$, $M_Y = 51.8 \, nT$ and $M_Z = -18.0 \, nT$ in GphiO coordinates (where X points in the direction of the incident Jovian magnetospheric flow, Y points along the Ganymede-Jupiter line and Z is parallel to the Jovian spin axis).
The Jovian magnetic field components were set to $B_X = 0 \, nT$, $B_Y = -6 \, nT$ and $B_Z = -77 \, nT$.
The inflow velocity was $V_X = 140 \, km/sec$, $V_Y = 0 \, km/sec$, $V_Z = 0 \, km/sec$.
The Jovian magnetospheric mass density and pressure were set to $\rho = 56 \, amu-cm^{-3}$ and $3.8 \, pPa$, respectively.

While BATS-R-US provides the option to run the Hall model with several non-ideal contributions included in Ohm's law, ${\bf E} = -{\bf V \times B}/c + {\bf J \times B}/(nec) - {\bf \nabla} p_e/(ne) + \eta {\bf J}$ (here, ${\bf V}$ is the plasma bulk velocity, ${\bf B}$ is the magnetic field, $c$ is the speed of light, ${\bf J}$ is the current density, $n$ is the plasma number density, $e$ is the electron charge, and $\eta$ is the plasma resistivity), the results presented here used the much simpler Ohm's law:  ${\bf E}_{HALL} \equiv -{\bf V \times B}/c + {\bf J \times B}/(nec)$.
Our neglect of the electron pressure gradient and resistivity terms was motivated by our desire to test, in the global magnetospheric context, the two primary results of the GEM Reconnection Challenge papers (summarized in \citet{birnc}):  1) ${\bf E}_{HALL}$ is the minimally complex Ohm's law required to produce reconnection rates comparable to those observed in full PIC simulations, 2) the reconnection rate is insensitive to the plasma resistivity when the ${\bf J \times B}/(nec)$ term is included in Ohm's law.
As we will see below, relying on numerical resistivity also allows us to more easily achieve a significant separation (factor of $\sim 5-10$) between $d_i$ and the dissipation scale.

The Hall and electron pressure terms introduce whistler and kinetic Alfv\'{e}n wave dynamics, respectively, both of which alter the structure of the reconnection layer in qualitatively similar ways (e.g., \citet{rogersa}).
Thus, by neglecting the pressure gradient term in Ohm's law, we are leaving out potentially important kinetic Alfv\'{e}n wave dynamics.
The kinetic Alfv\'{e}n wave physics, however, does not strongly influence the current sheet structure or reconnection rate for small guide fields; thus, neglecting it is appropriate for the G8 flyby case considered here.
Future simulations of other Galileo flybys where the reconnection guide field is moderately large will require inclusion of the electron pressure gradient term.
In any case, simply including the Hall term at least captures the de-coupling of electron and ion bulk velocities which is completely absent in resistive MHD, and this electron-ion decoupling below the ion inertial scale allows magnetic reconnection to occur on an Alfv\'{e}nic time scale that is comparable to that observed in resistive MHD with current-dependent anomalous resistivity.

Our Hall MHD model also neglects ion pressure anisotropies and gyro-viscous effects (e.g., the well-known ``gyro-viscous cancellation'' effect \citep{hazeltinea}  in the momentum equation).
While these effects -- and other finite Larmor radius (FLR) effects -- may also have important consequences for the structure of Ganymede's magnetosphere,  we will see below that the Hall term in Ohm's law already captures globally significant ion drifts that are absent in MHD .
Indeed, we argue below that many of the new effects observed in our Ganymede simulations (e.g., the asymmetric ``double magnetopause" boundary layer) have also appeared in multi-fluid simulations \citep{bennaa} and hybrid simulations \citep{mullera} of Mercury and can be explained (at least qualitatively) by two-fluid magnetic reconnection physics.

For the parameters described above, the Jovian magnetosphere ion inertial length was $d_i \approx 450 \, km = 0.17 \, R_G$.
Convergence experiments in which we compared results at increasingly higher resolutions demonstrated that accurately capturing the Hall effect within the magnetopause and magnetotail current sheets required at least $5$ computational cells per $d_i$; thus, for the results presented here, we chose a computational mesh consisting of several nested uniform grids with the innermost grid having a cell size of $\Delta X = \Delta Y = \Delta Z = 1/32 \, R_G \approx 83 \, km$ and dimensions $L_X = 7 R_G, L_Y = 4.5 R_G, L_z =4 R_G$ (see Figure \ref{FigGrid}).
This innermost grid was chosen to properly resolve the ion inertial scale in a region containing the upstream magnetopause and magnetotail current sheets.
The outer grid extended to $64 R_G$ to capture the Alfv\'{e}n wing structure far from the region of interest in the inner magnetosphere.

Our strategy was to focus on resolving the $d_i$ scale, relying on numerical resistivity $\eta$ to break the MHD frozen flux constraint.
Our reliance on numerical resistivity was made necessary by the fact that Hall effects do not become significant until the dissipation scale $l_\eta$ is a factor of $5-10$ smaller than $d_i$; however, resolving a resistive layer of thickness $l_\eta$ with $5-10$ cells when $l_\eta$ is itself a factor of $5-10$ smaller than $d_i$ would require $25-100$ computational cells per $d_i$, and this is currently still out of reach with present day computers.
Thus, relying on numerical resistivity with $l_\eta \propto \Delta X$ allowed us to accurately resolve $d_i$ scale structure with the minimal number of computational cells per $d_i$.
While this approach raises questions about how the results depend on the resistivity model, we point out that prodigious numerical evidence from Hall MHD simulations strongly suggests that the Hall-mediated reconnection rate is insensitive to the dissipation physics (e.g., see the GEM reconnection challenge papers \citep{birnc}).
Consistent with these past results, we have found in our convergence tests that the $d_i$ scale structures begin to converge when $\Delta X <~ d_i/5$, despite the fact that the resistivity is numerical.

We imposed inner boundary conditions on cells within a sphere of radius $1.0 \, R_G$.
The mass density and temperature at the inner boundary were initially set to $550 \, amu/cm^3$ and $20 \, eV$, respectively (consistent with \citet{jiaa}) and then allowed to float (with zero radial derivative) thereafter (in contrast to \citet{jiaa}, who fixed these values at the inner boundary).
While \citet{jiaa} surrounded their inner boundary with a conducting shell (to mimic the ionosphere) and imposed zero bulk velocity at the solid surface inside this conducting shell (thus allowing magnetospheric convection to drive ionospheric convection), we allowed the azimuthal components of the bulk velocity to float (zero radial derivative), while the radial component of the bulk velocity was chosen so that flow into the moon's surface was absorbed and flow out of the moon's surface was zero.
The magnetic field at the inner boundary was treated by splitting off Earth's dipole field and solving for the non-potential field $B_1$ (e.g., see \citet{tanakaa,powella}).
The total field was initialized to Ganymede's dipole field (i.e., $B_1 = 0$) and then $B_1$ was allowed to float with zero radial derivative thereafter.
Given the large uncertainties in the ionospheric conductivity -- and the complex nature of interaction of precipitating electrons and ions with Ganymede's atmosphere and icy surface (an interaction that involves both sputtering of neutrals and the production of secondary electrons from the surface) -- we believe this is a reasonable first step.
Future work will incorporate a more realistic model of the interaction of magnetospheric plasma with Ganymede's icy surface.

\begin{figure*}[th]
\centering
\includegraphics[width=6in,height=6in]{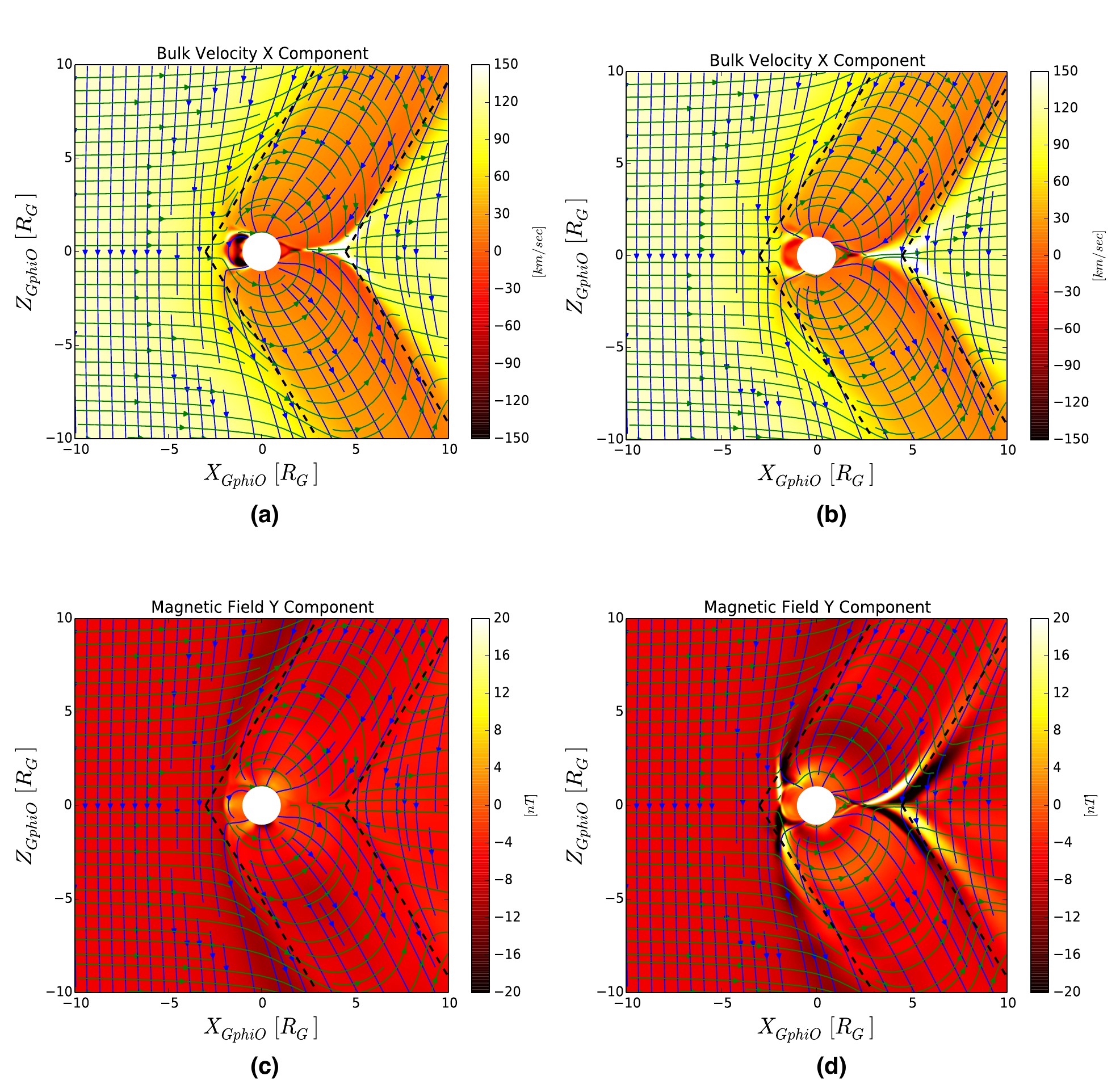}
\caption{\label{FigXZOverview} While the Hall effect does not have much impact on the large scale Alfv\'{e}n wing structure, it has a dramatic impact on the pattern of field-aligned currents near the moon.
Figures (a) and (c) show the resistive MHD case, while (b) and (d) show the Hall MHD case.
Streamlines of the magnetic field projected into the plane are shown in blue, and bulk velocity streamlines are shown in green.
The black dashed lines have slope $\pm \cos^{-1}({V_{in}/V_A})$; where $V_{in}$ is the ambient Jovian co-rotation velocity, and $V_A$ is the Alfv\'{e}n speed based on the ambient Jovian density and magnetic field.
The lower right panel shows the characteristic quadrupolar out-of-plane magnetic field pattern that characterizes Hall reconnection.
These out-of-plane fields are supported by field-aligned currents that merge into the Alfv\'{e}n wings and extend all the way down to the moon's surface.}
\end{figure*}

Before running the Hall MHD case, we ran a resistive MHD case with identical boundary conditions but using an anomalous resistivity of the following form:  $\eta = \eta_0 + \eta_A (J/J_C - 1)$, limited by $0$ and $\eta_{max}=2\times10^{10} m^{2}/s$ (here, $J$ is the magnitude of the current density; $\eta_0=1\times10^{9} m^{2}/s$, $\eta_A=2\times10^{9} m^{2}/s$ and $J_C=1\times10^{-7} A m^{-2}$ are constants)
As we will see below, our results for this case are comparable to those of \citet{jiaa} and thus serve as a good starting point from which to assess importance of the Hall effect.
We tested a variety of boundary conditions (not reported here), and found little impact on the final results. 

For the resistive MHD simulations, the BATS-R-US code was run in``local time stepping" mode (in which each grid cell is advanced at the largest time step possible for numerical stability) until a quasi-steady state was reached; then, ``time-accurate" time stepping was used.
For the Hall MHD runs, we used two different initial states:  1) local time stepping as for the MHD simulations, 2) the final time-accurate MHD state.
We ran the simulations in time-accurate mode long enough to reach a quasi-steady state ($\sim 10$ minutes of simulated time was found to be sufficient for Ganymede's small magnetosphere).
We obtained similar results in both cases, giving us confidence that the final Hall MHD state presented below was not influenced by the initial condition.

\section{Results}

\begin{figure*}[th]
\centering
\includegraphics[width=6in,height=2in]{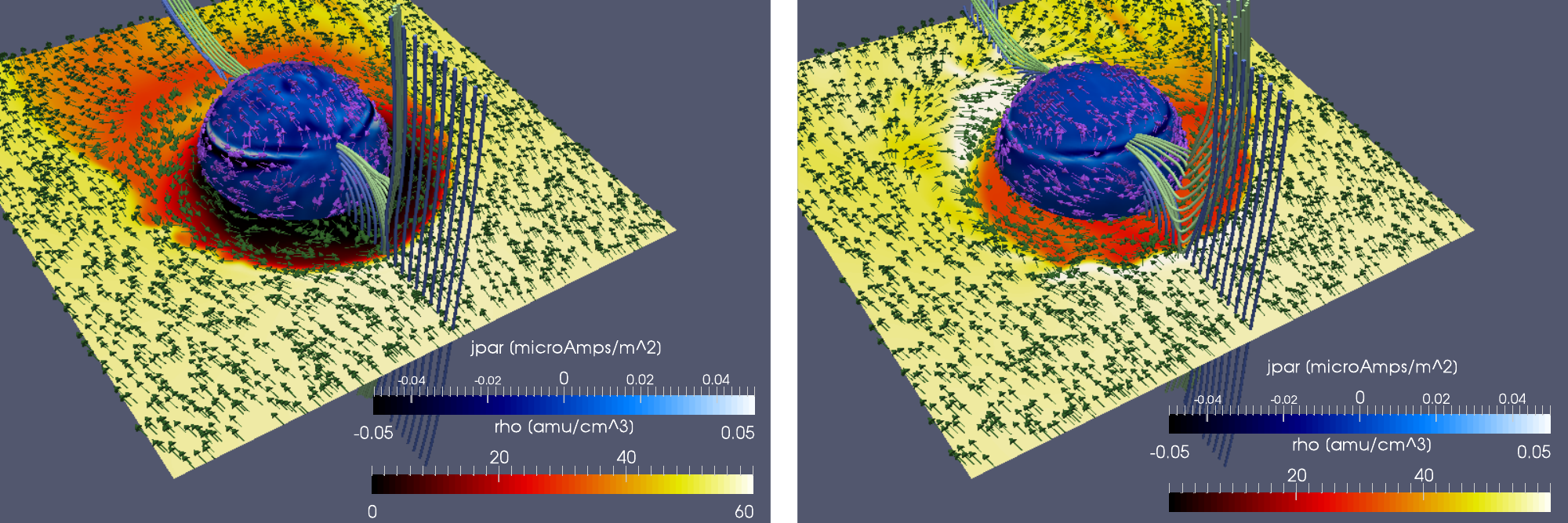}
\caption{\label{FigOverview3D} In resistive MHD (left panel), the magnetospheric convection and field-aligned currents (FAC) are symmetric about the Jovian magnetospheric inflow direction, exhibiting the classic Dungey convection (green arrows on the density plane and purple arrows on the inner sphere) and region-1 FAC (blue-white colorbar on the inner sphere) patterns.
In constract, the Hall simulation (right panel) shows strong asymmetries in both the convection and FAC patterns.
In the Hall simulation, Jovian magnetospheric plasma appears to enter Ganymede's magnetosphere along the sub-Jovian flank, populating the plasma sheet with Jovian magnetospheric plasma (in contrast to the resistive MHD case, in which the wake region is more or less devoid of Jovian magnetospheric plasma).
Note the pronounced bending of reconnected field lines (green tubes) out of the reconnection plane in the Hall case.}
\end{figure*}

Figure \ref{FigXZOverview} shows an overview of the Hall and resistive MHD simulation results.
The two upper figures show the $X$ (GphiO) component of the bulk velocity for the resistive (left) and Hall (right) runs, showing that the large scale Alfv\'{e}n wing structure is not significantly modified by the Hall effect.
Reconnection at the upstream magnetopause and downstream tail current sheets, however, produces the characteristic quadrupolar out-of-plane magnetic field pattern (lower right panel).
The out-of-plane field is supported by a system of intense field-aligned currents (FAC) that merge into the large scale Alfv\'{e}n wings far from the moon, but these currents also extend all the way down to the moon's surface.

Looking at the $X-Y$ plane (Figure \ref{FigOverview3D}), however, reveals that the Hall effect introduces significant asymmetries into the global convection and FAC patterns.
In resistive MHD, Jovian magnetospheric plasma flows around the upstream magnetopause symmetrically, producing a density cavity devoid of Jovian magnetospheric plasma in Ganymede's wake.
This is the classic Dungey \citep{dungeya,dungeyb} convection cycle, driven by magnetic reconnection at the upstream magnetopause and downstream tail current sheets.
Associated with this symmetric convection pattern is an antisymmetric pattern of FAC, most pronounced on the flanks and vanishing at the upstream and downstream edges of the polar cap.

In contrast, in the Hall MHD simulation Jovian magnetospheric plasma enters Ganymede's magnetosphere along the sub-Jovian flank, filling Ganymede's wake with Jovian magnetospheric plasma.
The bending of the reconnected field lines (green tubes) out of the reconnection plane -- which produces the classic quadrupolar out-of-plane magnetic field pattern shown in the lower right panel of Figure \ref{FigXZOverview} -- generates a new system of field-aligned currents that are most intense near the upstream and downstream edges of the polar cap.
The global convection pattern also shows large asymmetries, with Jovian magnetospheric plasma flowing into Ganymede's wake region through the sub-Jovian flank.
We demonstrate below that all of these effects are directly driven by Hall electric fields within the upstream magnetopause and downstream tail current sheets and, thus, should be present to some degree in all planetary magnetospheres (e.g., Mercury and Earth) for which convection is driven by collisionless magnetic reconnection.

\begin{figure*}[th]
\centering
\includegraphics[width=3in,height=4.5in]{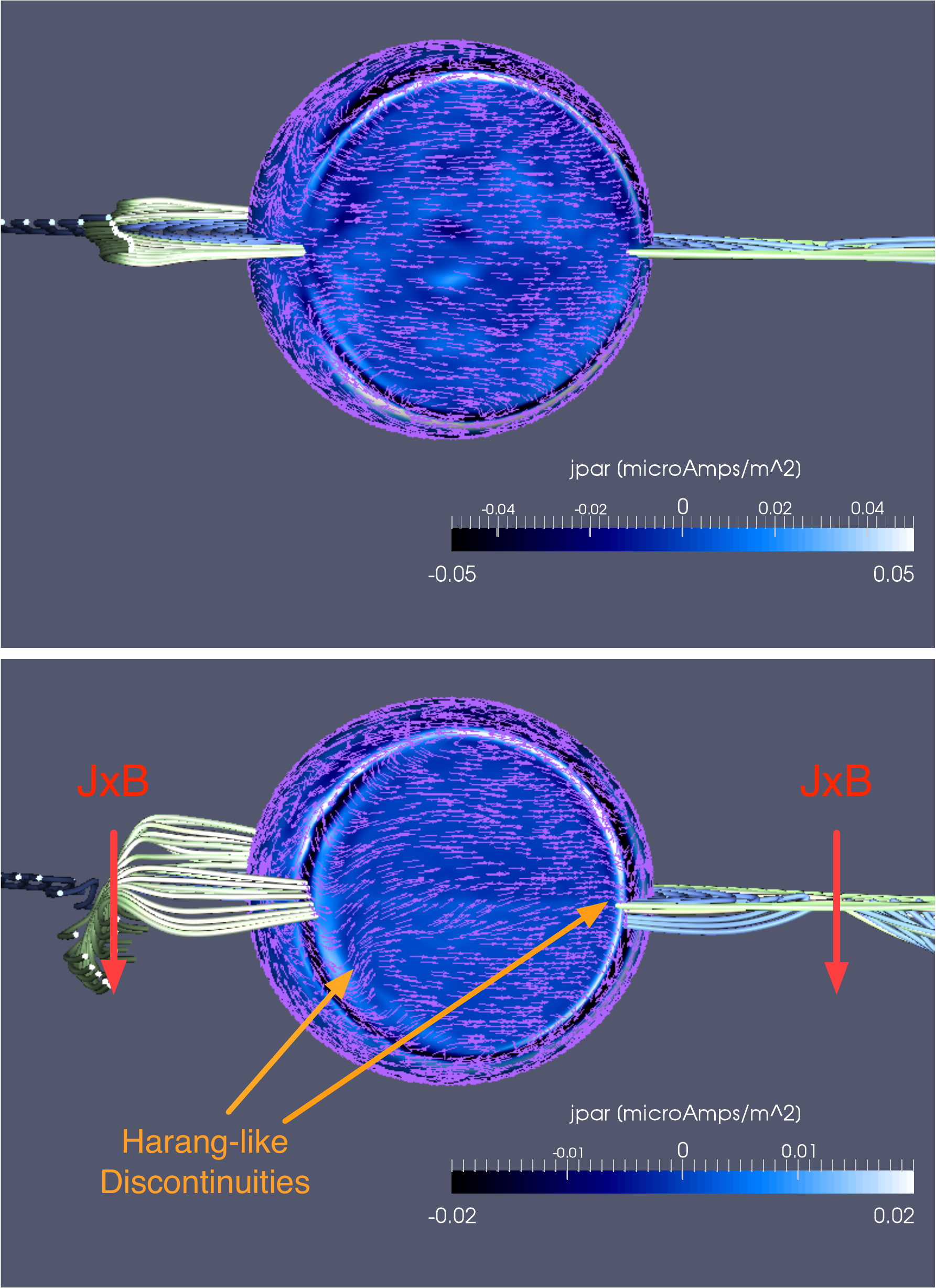}
\caption{\label{FigPolarCapFAC} The resistive MHD polar cap convection pattern shows the classic symmetric Dungey cycle, driven by magnetic reconnection at the upstream magnetopause and tail current sheets.
Plasma flows uniformly over the polar cap from the upstream side to the downstream side, and the FAC pattern shows the familiar relationship between flow shear (responsible for magnetic field line bending in the magnetosphere) and current density.
In the resistive MHD case, the current densities are most pronounced in the flanks.
In contrast, the Hall MHD convection pattern exhibits large asymmetries, driven by field line bending associated with Hall reconnection.
The field line bending produces new ${\bf J \times B}$ forces on the ions, generating new ``Harang-like" discontinuities in the convection pattern.}
\end{figure*}

Figure \ref{FigPolarCapFAC} shows a closer view of the FAC and convection patterns in the polar cap.
The resistive MHD case shows an anti-symmetric pattern of FAC into the moon's atmosphere, with parallel current densities of about $0.05 \mu-A/m^2$ (consistent with those reported by \citet{jiaa,jiab}).
In the Hall case, the FAC pattern is distorted by the reconnection FAC system, and now the most intense current densities appear in bands centered at the upstream and downstream edges of the polar cap.
The polar cap convection pattern in the Hall case shows localized acceleration channels near the upstream and downstream polar cap boundaries, both of which are associated with the bending of the magnetic field lines out of the reconnection planes in the magnetopause and magnetotail current sheets (labelled ``JxB" in the figure).
We will refer to these acceleration channels as ``Harang-like discontinuities," since they are similar in appearance to the Harang Discontinuity observed in the evening sector convection electric field of Earth's ionosphere.

The Hall reconnection physics is confined to $d_i$ scale layers and, as shown in Figure \ref{FigXZOverview}, has little impact on the large scale Alfv\'{e}n wing structure.
However, Hall electric fields appear to have a dramatic impact on the global convection pattern in the $X-Y$ plane.
Figure \ref{FigXYConvection} shows cuts of current density magnitude and bulk velocity streamlines (green lines and arrows) in the $X-Y$ plane for the resistive MHD run (left panel) and the Hall run (right panel).
The MHD case shows the classic Dungey convection pattern, with tail reconnection driving flow jets moonward and tailward from an X line oriented more or less along the $Y$ axis about $2 R_G$ downstream of the moon.
Note the symmetric pattern of Kelvin-Helmholtz (KH) waves on the flanks, driven by the flow shear between the reconnection return flow and the Jovian co-rotation inflow.

In the Hall case (right panel of Figure \ref{FigXYConvection}), large scale bulk flow channels appear, flowing in the $-Y$ direction within the magnetopause and magnetotail current sheets (magenta arrows in the figure).
These jets produce an asymmetric pattern of KH waves on the upstream magnetopause, with larger amplitude oscillations appearing on the sub-Jovian side of the magnetopause.
Qualitatively, the asymmetry can be explained by the asymmetry in the flow shear between the incoming Jovian co-rotation flow (blue arrows) and the upstream magnetopause current sheet jet (magenta arrow).
Jovian magnetospheric plasma enters Ganymede's magnetosphere through the sub-Jovian flank, returning along a thickented sub-Jovian boundary layer to join with the upstream magnetopause jet.
The resulting high flow shear on the sub-Jovian side of the magnetopause produces large amplitude KH waves there.

A similar effect explains the patchy appearance of the tail current density in the right panel of Figure \ref{FigXYConvection}:  KH waves driven by the tail ion jet cause the current sheet to undulate in and out of the $X-Y$ plane, producing the appearance of current dropouts in the plane.
Figure \ref{FigYZPlane} shows how these KH waves develop in the tail current sheet.
The arrow glyphs show the plasma bulk velocity, painted by magnitude, showing how plasma flowing into the tail current sheet interacts with the current sheet jet to produce large flow shear on the anti-Jovian side, where the KH wave amplitudes are largest.

\section{Comparison with Galileo magnetometer observations}

To test the model predictions described above, we probe our simulated magnetosphere with a virtual Galileo probe corresponding to the G8 flyby.
The G8 flyby crossed the upstream magnetopause at about $0.75 \, R_G$ above the $X-Y$ plane.
Since upstream magnetopause reconnection should be occurring continuously and more or less steadily (given the steady strongly southward JMF conditions), the G8 inbound magnetopause pass should have been an ideal opportunity for the Galileo magnetometer (see the overview by \citet{kivelsonb})  to observe the out-of-plane Hall electric field in the exhaust region above the reconnection site.
Not surprisingly, previous comparisons with resistive MHD by \citet{jiaa} did not see the expected Hall fields, but they did note the presence of large amplitude magnetic field fluctuations near the magnetopause crossings.
Later, \citet{jiac} used resistive MHD simulations to interpret the observed magnetic field fluctuations as the result of bursty reconnection producing large scale flux rope structures, or Flux Transfer Events (FTEs), similar to those observed at Earth's magnetopause (see the review by \citep{elphica}).
In what follows, we argue that these large scale magnetic field fluctuations on the inbound crossing are in fact steady state structures associated with collisionless reconnection.

\begin{figure*}[th]
\centering
\includegraphics[width=7in,height=3in]{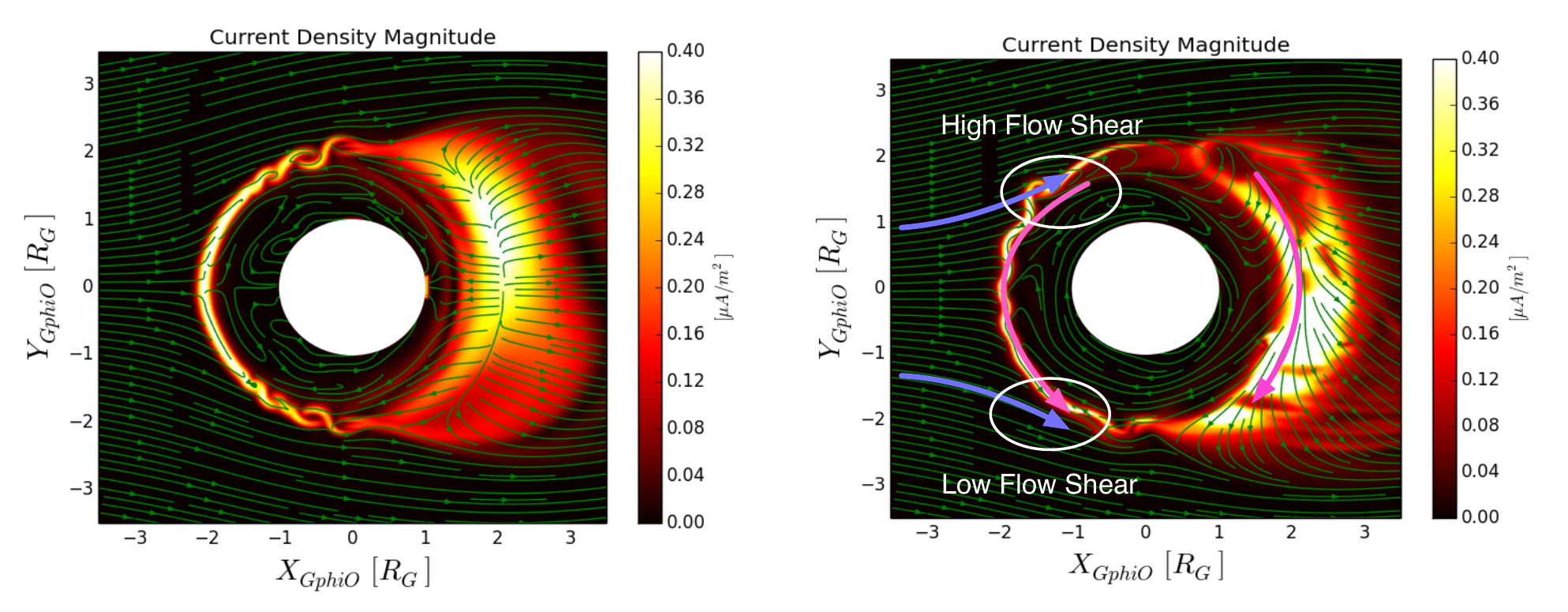}
\caption{\label{FigXYConvection} The Hall effect produces large asymmetries in Ganymede's convection pattern.
The left panel shows the classic Dungey convection pattern for the resistive MHD case, in which plasma is accelerated both moonward and tailward away from an X line more or less aligned with the $Y$ axis.
In the Hall case (right panel), fast ion jets within the upstream magnetopause and tail current sheets allow Jovian magnetospheric plasma to circulate throughout Ganymede's magnetosphere, entering from the sub-Jovian flank and returning along a thickened sub-Jovian boundary layer to join with the upstream magnetopause jet.
This asymmetric circulation pattern produces a corresponding asymmetric pattern of Kelvin-Helmholtz waves on the upstream magnetopause.}
\end{figure*}

\begin{figure*}[th]
\centering
\includegraphics[width=3in,height=3in]{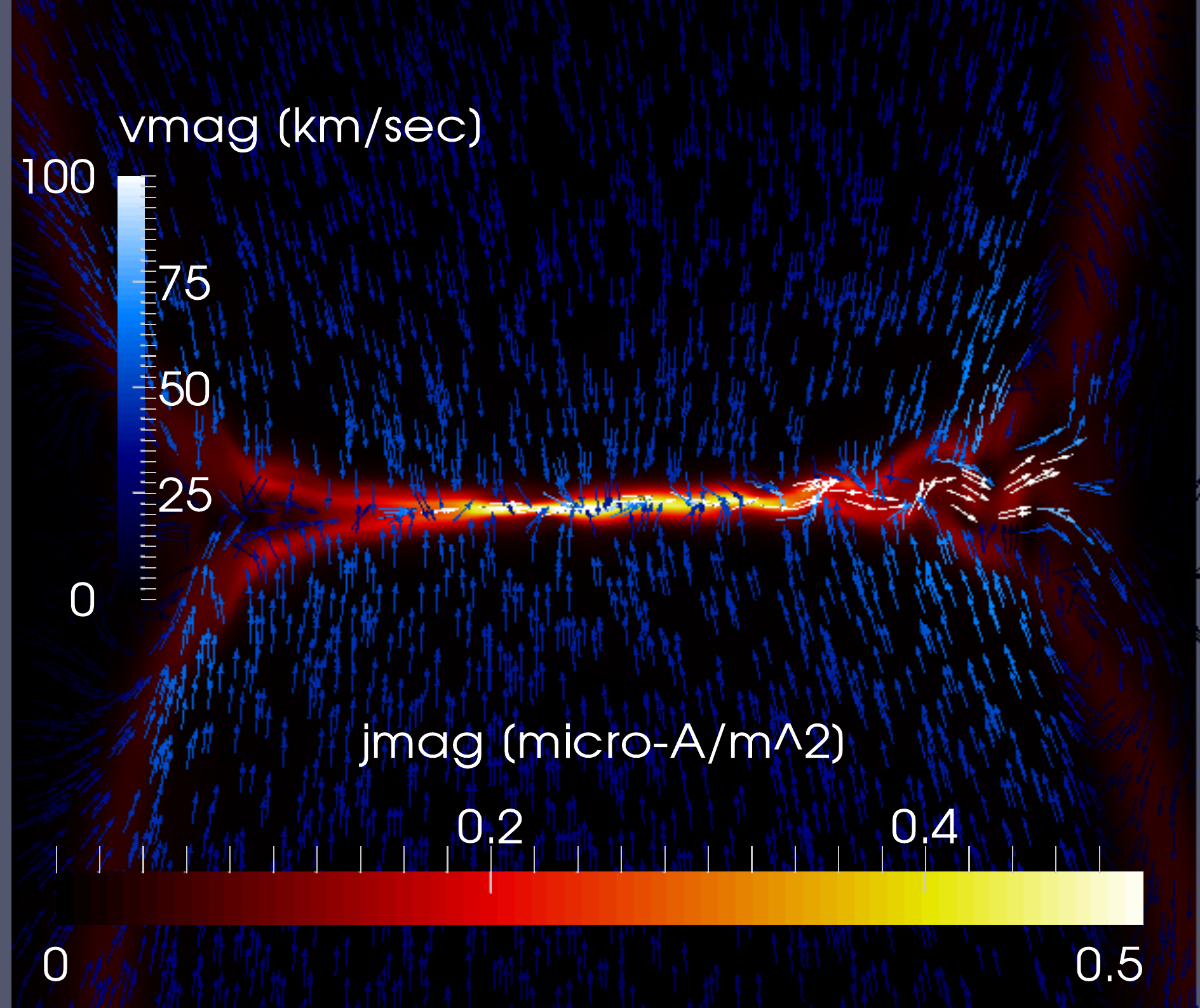}
\caption{\label{FigYZPlane} The Hall effect produces an ion jet in Ganymede's magnetotail current sheet.
This in turn produces a large flow shear on the anti-Jovian side of the current sheet (toward the right in the plot).
Here, we show a cut plane at $X = 2.5$ ($Y$ is horizontal and $Z$ is vertical), showing how the resulting Kelvin-Helmholtz (KH) waves cause an undulation of the current sheet that appears as small scale current density drop-outs in the right panel of Figure \ref{FigXYConvection}.
The arrows show bulk velocity color coded by magnitude.}
\end{figure*}

\begin{figure*}[th]
\centering
\includegraphics[width=6in,height=4.5in]{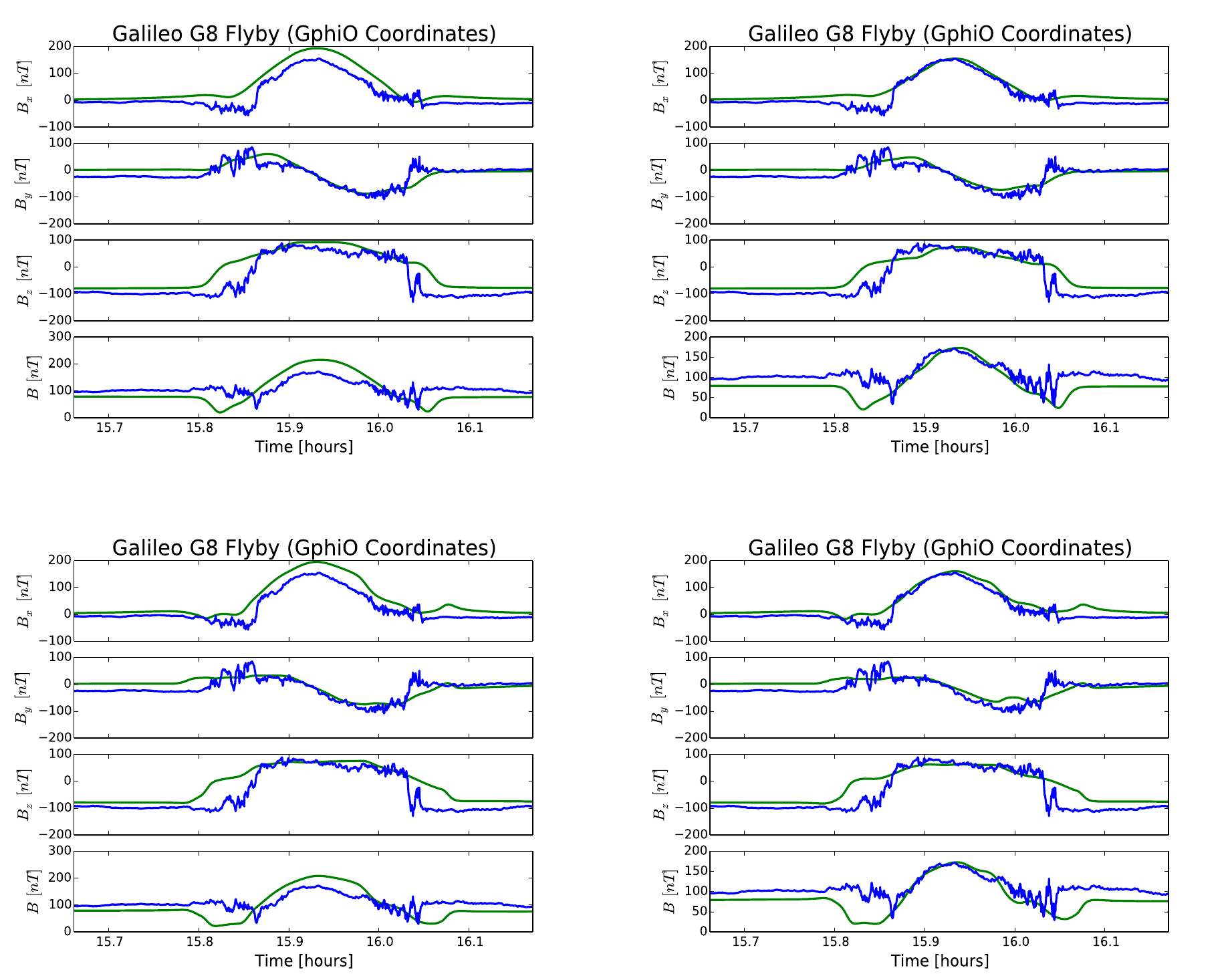}
\caption{\label{FigG8ProbeGphiO} Comparisons of our simulations with the Galileo magnetometer observations for the G8 flyby show good overall agreement.
The left panels show the comparison using the published G8 flyby coordinates.
Green lines show the simulation results, and blue lines show the data.
The right panels show the results of adding to the simulated probe an offset of $+0.05 \, R_G$ in the $X$ and $Z$ position components.
The top panels show the resistive MHD case, and the two bottom panels show the Hall MHD case.}
\end{figure*}

Figure \ref{FigG8ProbeGphiO} compares observed (blue lines) and simulated (green lines) magnetic field components along the G8 flyby.
The top two panels show the resistive MHD results, and the bottom two panels show the Hall results.
The left panels show the magnetic field components along the published G8 trajectory.
While there is good qualitative agreement, the simulated probe shows a somewhat larger $B_X$ than observed.
This is not surprising, since small differences between the simulated and actual magnetopause shapes may produce observable systematic differences in the magnetic field components in the closed field line region (which become distorted near the magnetopause).
To explore this effect, we introduced a small offset to the published G8 trajectory ($+0.05 \, R_G$ in the $X$ direction, and $+0.05 \, R_G$ in the $Z$ direction).
This offset was chosen to improve the agreement between simulated and observed $B_X$ components, and the offset trajectory will be used in all further discussion below.

Because of the G8 trajectory crosses the magnetopause at a finite $Y$ location, the reconnection plane is not aligned with the GphiO $X-Z$ plane.
Thus, the Hall out-of-plane field would be expected to produce large perturbations in both the $X$ and $Y$ components.
However, rotating into a ``Boundary Normal Coordinate" (BNC) system (where the new $X$ unit vector points along the magnetopause outward normal, the $Z$ unit vector points northward tangent to the magnetopause, and the $Y$ unit vector completes the right-handed system) would be expected to eliminate the perturbation in the new $X$ component, leaving only the new $Y$ (out of the local reconnection plane) component.
This is exactly what we observe, as we show now below.

\begin{figure*}[th]
\centering
\includegraphics[width=6in,height=4.5in]{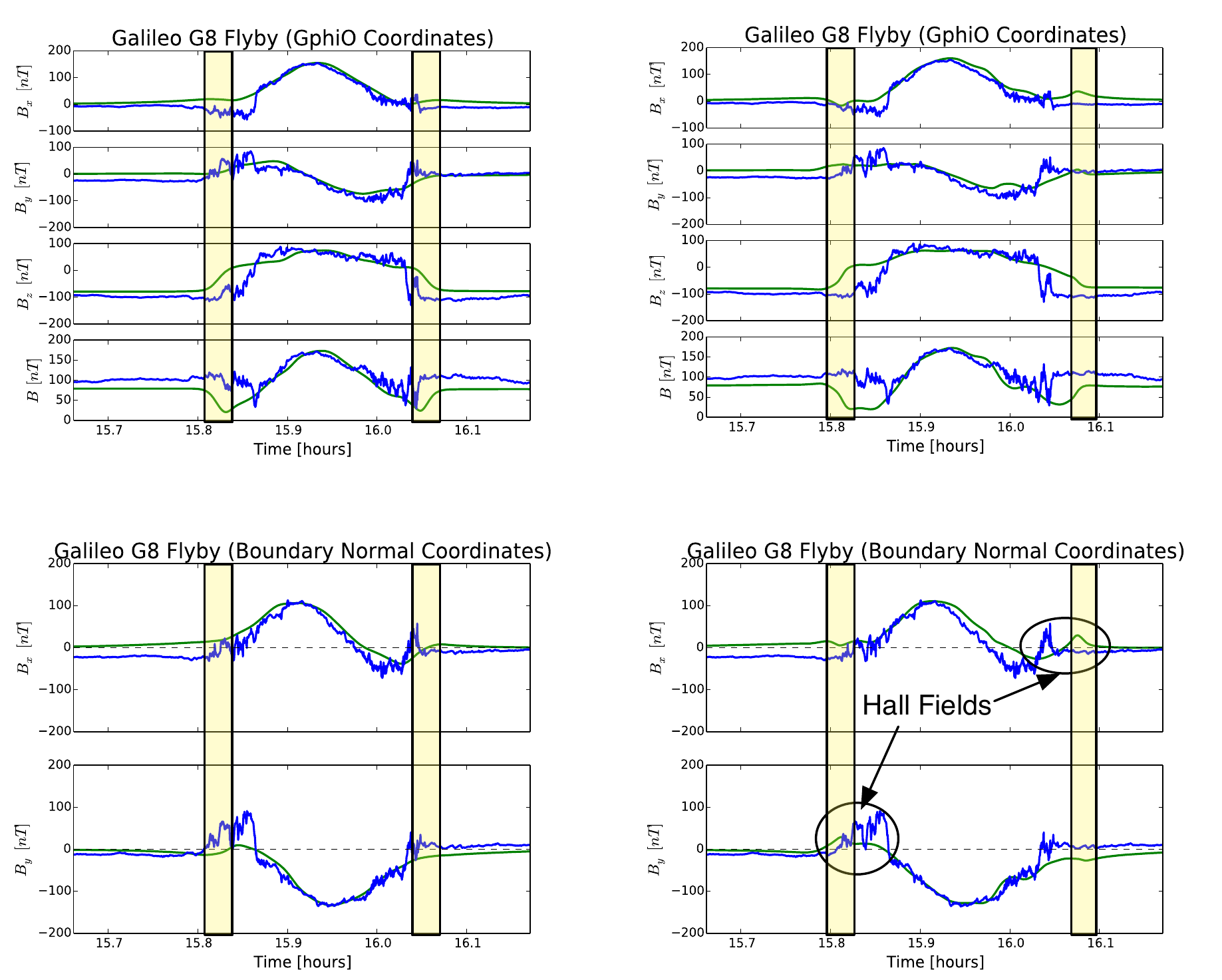}
\caption{\label{FigG8ProbeBNC} The Hall perturbations become particularly clear in Boundary Normal Coordinates (BNC).
Here, the left panels show the resistive MHD case, and the right panels show the Hall case, with yellow rectangles indicating the boundaries of the outer magnetopause current sheet in the simulations.
The perturbations marked ``Hall Fields" in the lower right panel show the out-of-plane magnetic field structure in the reconnection exhausts as they are crossed by the simulation probe.
Note that similar structures are observed in the actual magnetometer data.}
\end{figure*}

\begin{figure*}[th]
\centering
\includegraphics[width=3.5in,height = 5in]{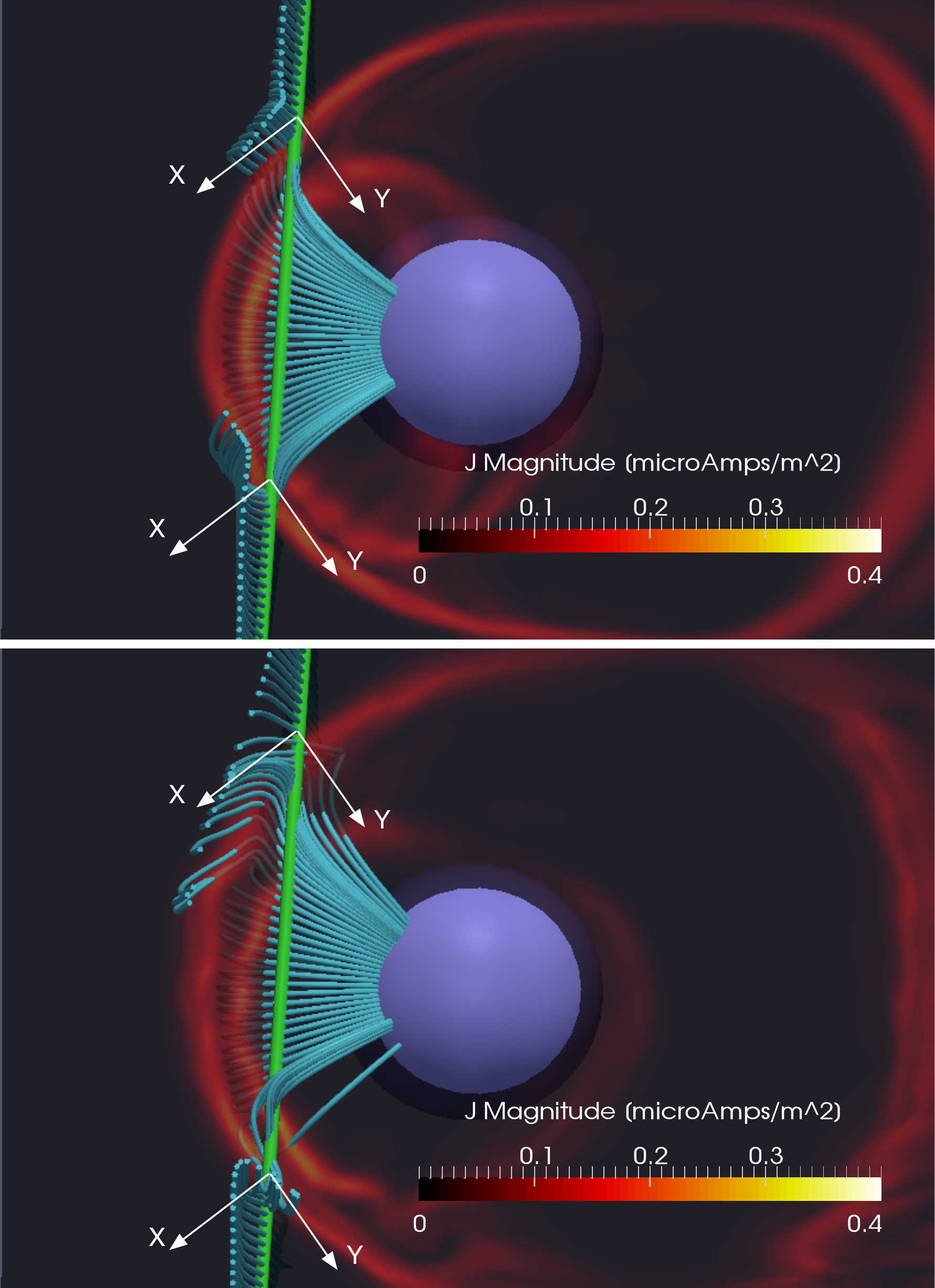}
\caption{\label{FigG8Probe3D} Simulations reveal that Ganymede's magnetopause has a bifurcated current sheet structure at the location of the G8 flyby (green line).
In the resistive MHD case, the current sheet is relatively smooth, and there are no large out-of-plane magnetic field excursions as the simulated probe crosses the boundary layer between the current sheets.
In the Hall case, however, the asymmetric pattern of Kelvin-Helmholtz (KH) waves is apparent, as are the large out-of-plane excursions of the magnetic field (blue field lines).
In both panels, the white $X-Y$ axes show the orientation of the inbound magnetopause crossing Boundary Normal Coordinate (BNC) system.}
\end{figure*}

Figure \ref{FigG8ProbeBNC} shows the data transformed into BNC coordinates corresponding to the inbound outer current sheet crossing, and Figure \ref{FigG8Probe3D} shows a three-dimensional view of the offset G8 flyby trajectory (green line) through our resistive (top panel) and Hall (bottom panel) simulations.
We determined the BNC normal direction by inspection of this plot for the inbound magnetopause crossing.
The corresponding coordinate axes are shown in white.
Note that the axis near the outbound (upper) crossing is simply translated from the inbound (lower) crossing, to simplify the interpretation of Figure \ref{FigG8ProbeBNC}, which plots the components in the single inbound current sheet BNC coordinate system.

The yellow rectangles in Figure \ref{FigG8ProbeBNC} show where the simulated probe enters and exits the outer magnetopause current sheet, which appears as an obvious reversal in the $B_Z$ component.
The second current sheet crossing is less obvious in both the simulations and the data and, therefore, we have not attempted to identify it in Figure \ref{FigG8ProbeBNC} (though it is obvious in Figure \ref{FigG8Probe3D}).
For reference, the GphiO magnetic field components are reproduced in the top two panels, with resistive MHD on the left and Hall MHD on the right.
As expected, transforming the data into inbound magnetopause BNC coordinates eliminates the negative $B_X$ excursion, leaving only an enhanced $B_Y$ excursion.
This is consistent with what we observe in the Hall simulation (right panels of Figure \ref{FigG8ProbeBNC}):  The small negative $B_X$ perturbation in the top right panel at about $15.8 hours$ has been transformed away in the bottom right panel, and the small positive $B_Y$ excursion in the top right panel has been enhanced in the lower right panel.
Similarly, when viewed in the inbound BNC coordinate system, a large positive $B_X$ perturbation appears on the outbound crossing.
We have labelled these perturbations ``Hall Fields'' in Figure \ref{FigG8ProbeBNC}.

While these features are somewhat difficult to see in the line plots of Figure \ref{FigG8ProbeBNC}, they appear very clearly in the 3D view of Figure \ref{FigG8Probe3D}.
The resistive MHD case in the top panel of Figure \ref{FigG8Probe3D} shows that there is no field line bending out of the reconnection plane as the simulated probe crosses the magnetopause boundary layer.
In contrast, Hall case in the bottom panel shows clearly that there is significant bending out of the reconnection planes of both the inbound and outbound crossings.
For the inbound crossing, it is obvious that the perturbed field is predominantly in the BNC $Y$ direciton, while the perturbed field is in the $X$ direction for the outbound crossing when viewed in the local BNC coordinate system for the inbound crossing.
This is exactly what is observed in the data.

Another interesting feature of the G8 Galileo magnetometer data is the series of large amplitude waves that appear (most clearly in the magnetic field magnitude) as Galileo is exiting Ganymede's magnetosphere on the sub-Jovian side.
These waves do not appear in the MHD simulations, but they do appear in the Hall simulations (compare the upper left and upper right magnetic field  magnitude traces in Figure \ref{FigG8ProbeBNC}).
Figure \ref{FigG8Probe3D} shows that the wave-like feature in the simulated outbound crossing is caused by the KH waves that are driven by enhanced flow shear on the sub-Jovian side of the upstream magnetopause (as argued above in the context of Figure \ref{FigXYConvection}).
To our knowledge, these large amplitude wave-like perturbations observed on the G8 outbound magnetopause crossing have not yet been interpreted as KH waves; our interpretation is that these KH waves -- appearing on the sub-Jovian side but not on the anti-Jovian side -- are a new feature predicted by our Hall model.
They are driven by the upstream magnetopause ion jet which, in turn, owes its existence to ion-scale structure of the reconnection site.
Thus, we argue that intermittent upstream magnetopause reconnection (and associated flux ropes or FTEs) is not required to explain the large amplitude magnetic field perturbations observed on the G8 inbound and outbound magnetopause crossings.
These structures can be explained as consequences of steady collisionless reconnection:  1) the classic out-of-plane magnetic field bending, and 2) the ion drift that supports the upstream magnetopause current sheet.
Neither of these effects is present in resistive MHD simulations.

\section{Physics of the ion drift belts}

While the appearance of the out-of-plane magnetic field is an expected and well-understood consequence of the Hall effect, the dramatic distortion of Ganymede's global convection pattern was a surprise to us.
While the sources of the new ion jets are confined to the upstream and tail current sheets, they clearly have a global impact on the structure of Ganymede's magentosphere.
The most obvious effect is the appearance of a new global system of ``ion drift belts" that help circulate Jovian magnetospheric plasma througout Ganymede's magnetosphere.
The most striking difference between the classic Dungey resistive MHD pattern and the new drift belt pattern is that Ganymede's wake should be filled with Jovian magnetospheric plasma in a thin (ion inertial scale) layer around $Z = 0$.
Another important global effect is the appearance of a significantly thickened sub-Jovian boundary layer.

\begin{figure*}[th]
\centering
\includegraphics[width=6in,height=3in]{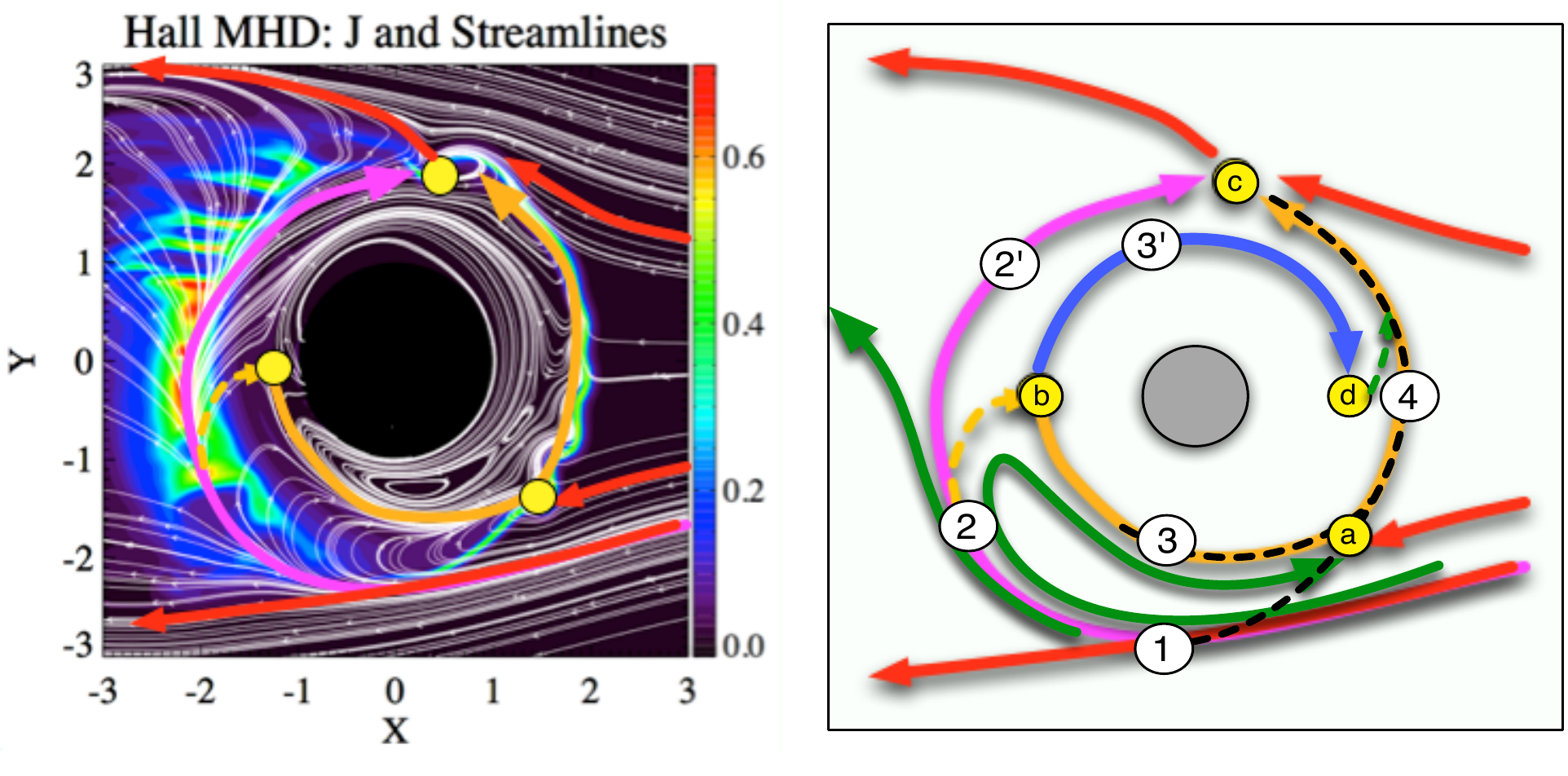}
\caption{\label{FigConvectionCartoon} The current sheet ion jets produce a global system of ``ion drift belts" that circulate Jovian magnetospheric plasma throughout Ganymede's magnetosphere.
The left panel shows the current density magnitude (colors) and flow streamlines (white lines) in the $X-Y$ plane (in a coordinate system for which $X$ points opposite to the Jovian co-rotation flow, $Z$ points along the Jovian rotation axis and $Y$ completes the right-handed system).
The right panel shows the path of a fluid particle as it circulates around the Jovian magnetosphere.
After entering Ganymede's tail along the sub-Jovian flank (1), the particle path splits into two paths, one of which (2,2') merges with the tail ion jet while another (3) circulates back around to the upstream side to merge with the upstream magnetopause jet (4).
This return path (1-2-3-4) results in the formation of a thickened sub-Jovian shear layer (green arrow) that produces a ``double magnetopause" structure (dashed black lines).
The returning particl path (3) also collides with the incoming Jovian magnetospheric plasma at stagnation point (a) to produce the sub-Jovian Kelvin-Helmholtz waves apparent in the left panel.}
\end{figure*}

Figure \ref{FigConvectionCartoon} illustrates our new ion drift belts model of global convection in Ganymede's magnetosphere.
The figure shows another view of the $X-Y$ cut shown in Figure \ref{FigXYConvection}, this time annotated with a cartoon of the flow topology.
The red arrows show the Jovian magnetospheric flow; the orange and magenta arrows show the upstream magnetopause and tail current sheet ion jets, respectively; the green arrows show how plasma of Jovian magnetospheric origin circulates through Ganymede's wake region and then back upstream, forming a thickened sub-Jovian ``double magnetopause" structure (note that Jupiter is somewhere below the bottom of the figure, co-rotating counter-clockwise).
The yellow circles show stagnation points of the flow.
As a parcel of Jovian magnetospheric plasma enters Ganymede's wake region on the sub-Jovian flank (1), it splits, with one path merging with the tail ion jet (2, 2') and the other path circulating back upstream (3) where it merges with the upstream magnetopause jet and collides with the incoming Jovian plasma (a) (generating KH waves in the process).
Meanwhile, the upstream jet (4) collides with the tail jet (2') at stagnation point (c'), where more KH waves are produced on the anti-Jovian flank.

\begin{figure*}[th]
\centering
\includegraphics[width=6in,height=2.5in]{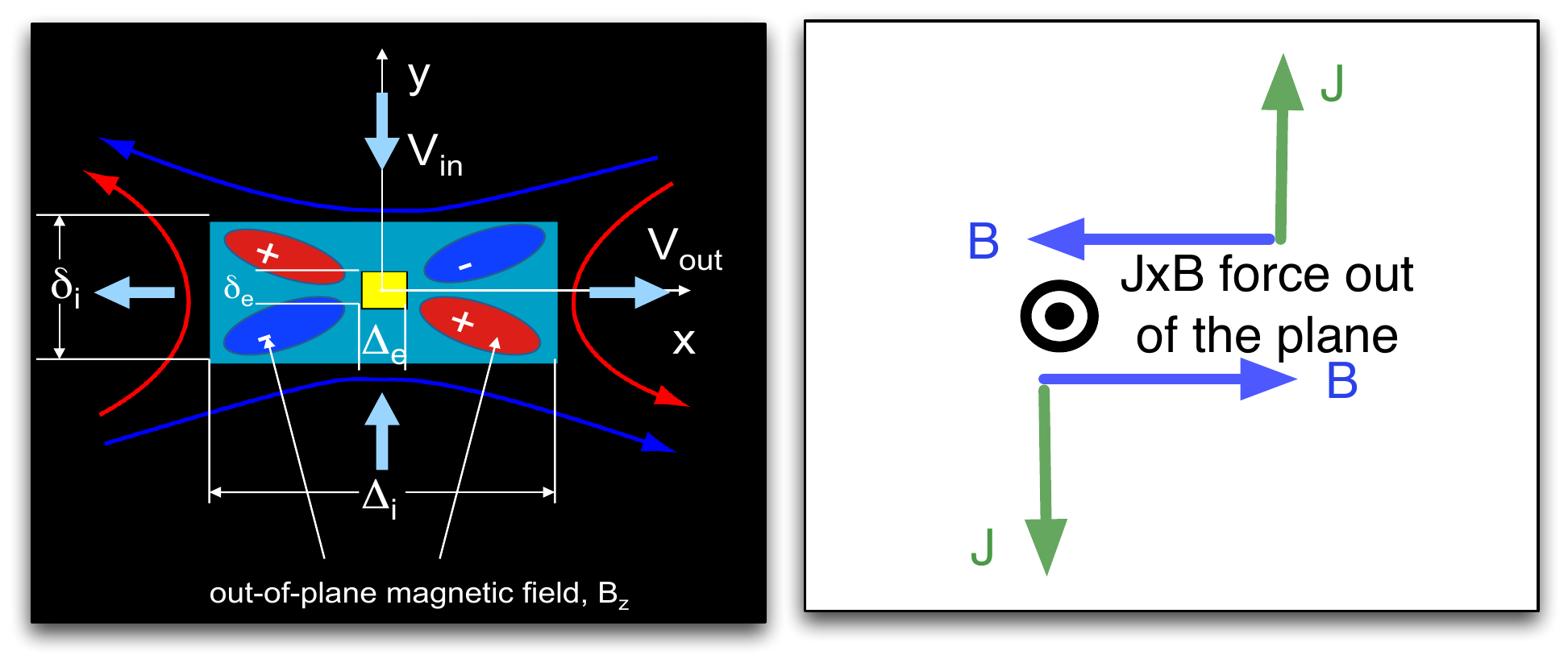}
\caption{\label{FigIonDriftPhysics} The acceleration mechanism responsible for the current sheet ion jets is a simple consequence of the Hall fields within the ion diffusion regions.
In the left panel, the cyan box is the ion diffusion region, and the yellow box is the electron diffusion region.
The blue and red ovals show the quadrupolar out-of-plane magnetic field pattern, supported by the in-plane Hall currents.
These in-plane currents produce an out-of-plane $J \times B$ acceleration, shown schematically in the right panel.}
\end{figure*}

The appearance of the Hall-induced ion drift belts is an easily understood consequence of local force balance near the reconnection X lines.
Figure \ref{FigIonDriftPhysics} illustrates the physics of the ion out-of-plane acceleration due to Hall currents in the reconnection plane.
In the left panel, the cyan box is the ion diffusion region -- the region over which the ion and electron bulk velocities become de-coupled.
In this region, electrons lead the ions into the electron diffusion region (yellow box), producing an in-plane current in the positive $Y$ direction above the $X$ axis and in the negative $Y$ direction below the $X$ axis (here, $X$ and $Y$ now refer to the outflow and inflow directions in the reconnection plane, respectively).
Similarly, electrons lead the ions out of the electron diffusion region, producing an in-plane current  in the negative $X$ direction to the right of the $Y$ axis and in the positive $X$ direction to the left of the $Y$ axis.
These in-plane current loops support the quadrupolar out-of-plane current system shown schematically by the red and blue ovals, and the resulting $J \times B$ force accelerate ions out of the reconnection plane.
This Hall acceleration mechanism illustrated schematically in the right panel.

We can estimate the speed of the upstream magnetopause ion jet by writing down the $Y$ (GphiO) component of the momentum equation:

\begin{equation}
\rho V_X {\partial V_Y \over \partial X} \approx - {J_X B_Z \over c}
\label{EqMomentumY}
\end{equation}

\noindent where the Hall current is given by Amp\'{e}re's law:  $J_X \approx -(c/4 \pi) \partial B_Y / \partial Z$, and $B_Y$ is the amplitude of the out-of-plane quadrupole field structure.
We approximate equation (\ref{EqMomentumY}) as follows:

\begin{equation}
\rho {V_X V_Y \over \delta_i} \approx {B_Y B_Z \over {\Delta_i 4 \pi}}
\label{EqMomentumYApprox}
\end{equation}

\noindent where $\delta_i$ and $\Delta_i$ are the thickness and length, respectively, of the ion diffusion region.
Assuming $\rho$ is approximately constant, mass conservation implies $V_X \Delta_i \approx V_Z \delta_i \approx V_A \delta_i$, where $V_A \equiv B_Z/(4 \pi \rho)^{1/2}$ (since ions leave the ion diffusion region at the upstream Alfv\'{e}n speed).
Further assuming that the amplitude of the Hall-induced quadrupole field is comparable to the Z component of the field upstream of the current sheet (i.e., $B_Y \approx B_Z$, as observed in the simulations) shows that the Hall-induced $J \times B$ force accelerates ions in the $Y$ direction to approximately the upstream Alvf\'{e}n speed as they cross the magnetopause current sheet:  $V_Y \approx V_A$.

\begin{figure}
\centering
\includegraphics[width=3.2in,height=2.8in]{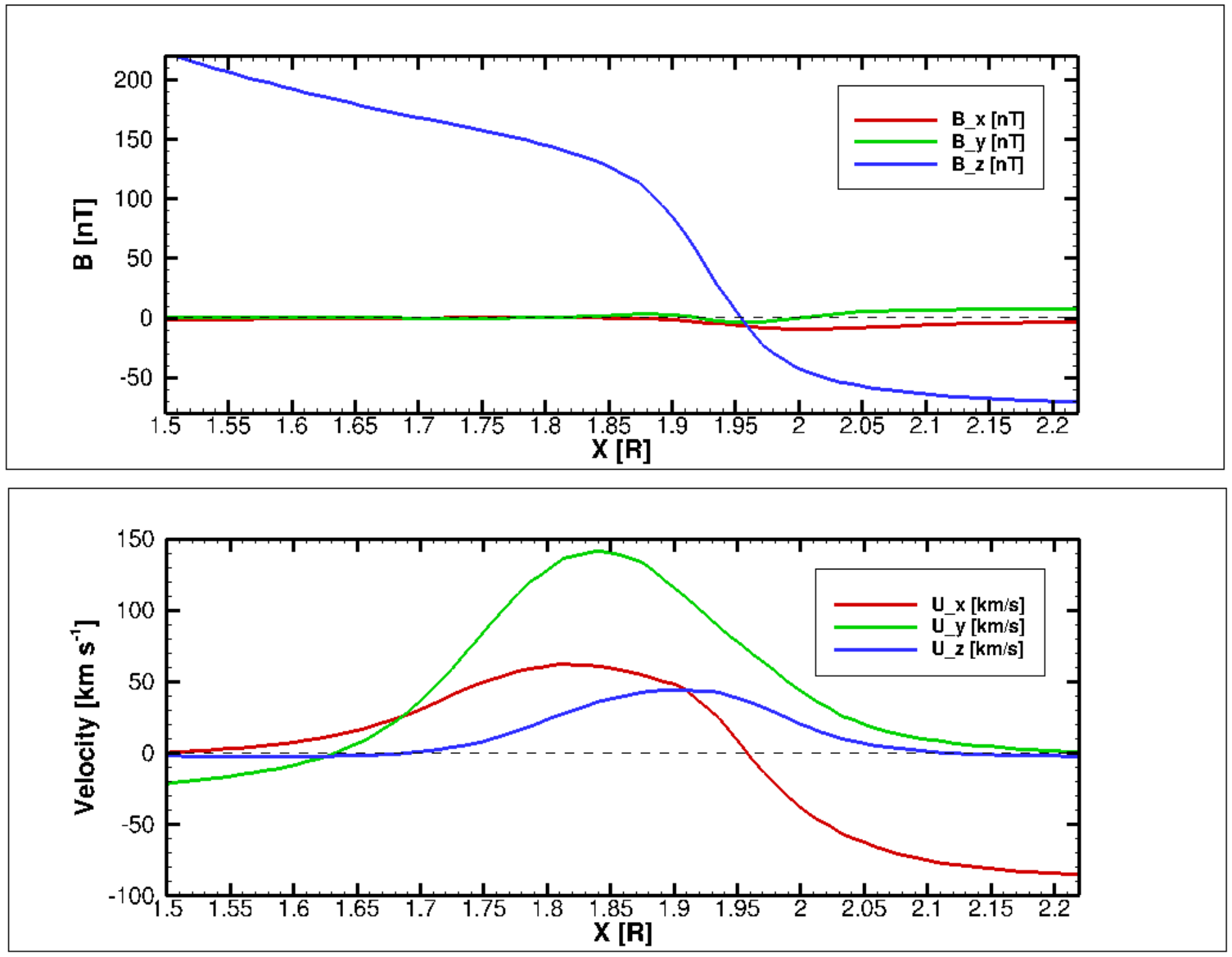}
\caption{\label{FigIonDriftCut} A cut along the $X$ axis crossing the upstream magnetopause confirms that the ions are accelerated to the local Alfv\'{e}n speed (in this case, $\approx \, 140 \,  km/sec$) in the out-of-plane direction.
The upper panel shows the three components of the magnetic field across the upstream magnetopause; the bottom panel shows the three components of the bulk velocity.}
\end{figure}

Figure \ref{FigIonDriftCut} shows a cut of the $Y$ component of the center-of-mass bulk velocity (green curve in the bottom panel) across the upstream magnetopause near $Y = Z = 0$.
Using the value $B_Z \approx 50 \, nT$ just upstream of the magnetopause (see the blue curve in the top panel), and the upstream mass density of $56 \, amu/cm^3$, we get an upstream Alfv\'{e}n speed of $V_A \approx 146 \, km/sec$, which compares favorably with the observed peak $V_Y$.
Thus, we are confident that the Hall effect explains the magnetopause ion jet.
Similar analysis also also confirms that the tail jet owes its origin to the Hall-induced $J \times B$ force.

\section{Conclusions and Discussion}

We have used high resolution Hall MHD simulations of Ganymede's magnetosphere to demonstrate that Hall currents within the magnetopause and magnetotail current sheets have a significant impact on the global structure of Ganymede's magnetosphere.
Specifically, the Hall effect introduces large asymmetries into the field-aligned current (FAC) and convection patterns:

\begin{enumerate}
\item Hall-mediated magnetic reconnection produces a new system of field-aligned currents that originate in the reconnection layers and extend all the way to the moon's surface.
\item Hall-mediated reconnection accelerates ions in the magnetopause and magnetotail to the local Alfv\'{e}n speed in the $Y$ direction, producing a global system of ion drift belts that circulate Jovian magnetospheric plasma throughout Ganymede's magnetosphere.
\end{enumerate}

The appearance of the new system of FAC, with maximum intensities near the upstream and downstream edges of the open-closed boundary, is interesting in light of the fact that a similar pattern is observed in Hubble observations (e.g., \citet{feldmanb} and \citet{mcgratha}).
In contrast, in resistive MHD, the region-1 type FAC actually vanish at the upstream and downstream edges of the polar cap and are enhanced on the flanks.
This is difficult to reconcile with observations if Ganymede's aurora is driven by field-aligned currents (as it is in Earth's magnetosphere).
While it is premature to claim that the Hall FAC system can explain Ganymede's aurora, it is encouraging that collisionless reconnection can serve as a steady magnetospheric generator for FAC in the right locations to explain the Hubble emission patterns.
Future work should quantify this, but such an effort would require a kinetic model of the interaction of the precipitating electrons and ions generated in the magnetosphere with Ganymede's atmosphere and icy surface (both of which may influence Ganymede's atmosphere).

Some observable consequences of the ion drift belts include:  a) large out-of-plane magnetic field perturbations associated with the Hall effect in the upstream magnetopause and downstream tail current sheets, b) an Alfv\'{e}nic ion jet extending across the low latitude upstream magnetopause, c) a thickened sub-Jovian magnetopause boundary layer with distinct outer and inner boundaries, d) an asymmetric pattern of magnetopause Kelvin-Helmholtz waves driven by the interaction of the incoming Jovian magnetospheric plasma with the upstream magnetopause ion jet, e) an Alfv\'{e}nic ion jet consisting of Jovian magnetospheric plasma extending across the central wake region, f) Harang-like discontinuities in the ionospheric convection pattern, and g) the presence of Jovian magnetospheric plasma in a thin $d_i$ scale layer (centered on $Z = 0$ close to the moon in the wake region.
We have argued in this work that some of these consequences (the large out-of-plane magnetic field perturbations and the asymmetric pattern of Kelvin-Helmholtz waves) are already apparent in the G8 magnetometer observations.
Some of the other consequences should be testable using the other Galileo flybys and including other Galileo instruments (plasma and waves), but such an investigation is beyond the scope of the present paper.
More extensive comparisons between simulations of other flybys and Galileo data will be pursued in future work.

We note that multi-fluid simulations by \citet{bennaa} have produced ion drift belts very similar to those found in our Ganymede simulations.
However, while \citet{bennaa} invoke electron pressure gradient drifts to explain the ion belts, our simulations did not include the electron pressure gradient in Ohm's law.
The similarity of our results to those of \citet{bennaa} suggest that Hall $J \times B$ forces can explain the ion drift belts.
We further speculate that the ``double magnetopause" structure observed in our Ganymede simulations is the same structure observed at Mercury by \citet{slavina} in the MESSENGER data:  a thickened dusk-side magnetopause boundary layer driven by Hall-mediated reconnection (rather than pressure gradient drifts, as argued by \citet{mullera}).
Further simulation work that isolates the Hall $J \times B$ force from pressure gradient drifts is required to definitively answer this question.

Since the effects observed in our simulations are directly driven by magnetic reconnection, they should be present in all magnetospheres.
The large scale consequences and observable signatures, however, will depend on a number of factors:  a) the size of the intrinsic magnetic field (which influences the size of the magnetosphere), b) the role of magnetic reconnection in driving global magnetospheric convection, c) the nature of the magnetized body's atmosphere and surface (which influences how field-aligned currents close).
We expect the Hall effect to have global influence in small magnetospheres for which convection is driven primarily by magnetic reconnection.
For example, the Hall effect likely plays a negligible role in the large scale structure of Jupiter's co-rotating magnetosphere.
Mercury, on the other hand, possesses a very small magnetosphere with convection driven by magnetic reconnection, so that Hall-mediated reconnection should produce global field-aligned current and ion drift belt structures.
Mercury's surface conductivity is very small, however, and it is not clear how the Hall-induced field-aligned currents will close.
Earth lies somewhere in between:  its magnetosphere is relatively large (several orders of magnitude relative to the ion inertial length), and its convection pattern transitions from an outer reconnection dominated region to an inner co-rotation dominated region.
Understanding the role of ion-scale reconnection in global magnetospheric structure is thus a compelling problem that will benefit greatly from future comparative simulation studies.

\begin{acknowledgments}
Input files used in the generation of the SWMF/BATS-R-US data files used in this study are available upon request to John C. Dorelli (john.dorelli@nasa.gov).
SWMF/BATS-R-US source code is available upon request to the Center for Space Environment Modeling (CSEM) (http://csem.engin.umich.edu/tools/swmf/downloads.php).
The SWMF license does not permit the authors to distribute the source code themselves.
The Galileo magnetometer data files used in this study are available upon request to John C. Dorelli (john.dorelli@nasa.gov).
\end{acknowledgments}

\bibliography{HallGanymedeJGR_v01}

\begin{thebibliography}{43}
\providecommand{\natexlab}[1]{#1}
\expandafter\ifx\csname urlstyle\endcsname\relax
  \providecommand{\doi}[1]{doi:\discretionary{}{}{}#1}\else
  \providecommand{\doi}{doi:\discretionary{}{}{}\begingroup
  \urlstyle{rm}\Url}\fi

\bibitem[{\textit{Benna et~al.}(2010)}]{bennaa}
Benna, M., et~al., Modeling of the magnetosphere of mercury at the time of the
  first messenger flyby, \textit{Icarus}, \textit{209}, 3--10, 2010.

\bibitem[{\textit{Birn et~al.}(2001)}]{birnc}
Birn, J., et~al., Geospace environmental modeling (gem) magnetic reconnection
  challenge, \textit{J. Geophys. Res.}, \textit{106}, 3715--3719, 2001.

\bibitem[{\textit{Biskamp et~al.}(1995)\textit{Biskamp, Schwarz, and
  Drake}}]{biskampc}
Biskamp, D., E.~Schwarz, and J.~F. Drake, Ion-controlled collisionless magnetic
  reconnection, \textit{Phys. Rev. Lett.}, \textit{75}, 3850--3853, 1995.

\bibitem[{\textit{Biskamp et~al.}(1997)\textit{Biskamp, Schwartz, and
  Drake}}]{biskampa}
Biskamp, D., E.~Schwartz, and J.~F. Drake, Two-fluid theory of collisionless
  magnetic reconnection, \textit{Phys. Plasmas}, \textit{4}, 1002--1009, 1997.

\bibitem[{\textit{Daughton et~al.}(2011)\textit{Daughton, Roytershteyn,
  Karimabadi, Yin, Albright, Bergen, and Bowers}}]{daughtond}
Daughton, W., V.~Roytershteyn, H.~Karimabadi, L.~Yin, B.~J. Albright,
  B.~Bergen, and K.~J. Bowers, The role of electron physics in the development
  of turbulent magnetic reconnection in collisionless plasmas, \textit{Nature
  Physics}, \textit{7}, 539--542, 2011.

\bibitem[{\textit{Dorelli}(2003)}]{dorellid}
Dorelli, J., Effects of hall electric fields on the saturation of forced
  antiparallel magnetic field merging, \textit{Phys. Plasmas}, \textit{10},
  3309--3314, 2003.

\bibitem[{\textit{Dorelli and Birn}(2003)}]{dorellic}
Dorelli, J., and J.~Birn, Whistler-mediated magnetic reconnection in large
  systems: Flux pile-up and the formation of thin current sheets, \textit{J.
  Geophys. Res.}, \textit{108}, 1133, 2003.

\bibitem[{\textit{Dungey}(1961)}]{dungeya}
Dungey, J.~W., Interplanetary magnetic field and the auroral zones,
  \textit{Phys. Rev. Lett.}, \textit{6}, 47--48, 1961.

\bibitem[{\textit{Dungey}(1963)}]{dungeyb}
Dungey, J.~W., The structure of the exosphere, or adventures in velocity space,
  in \textit{Geophysics, The Earth's Environment}, edited by C.~DeWitt,
  J.~Hieblot, and A.~Lebeau, p. 505, Gordon and Breach, New York, 1963.

\bibitem[{\textit{Elphic}(1995)}]{elphica}
Elphic, R.~C., Observations of flux transfer events: A review, in
  \textit{Physics of the Magnetopause}, \textit{AGU Monogr. Ser.}, vol.~90,
  edited by P.~Song, B.~U.~O. Sonnerup, and M.~Thomsen, pp. 225--233, American
  Geophysical Union, 1995.

\bibitem[{\textit{Feldman et~al.}(2000)\textit{Feldman, McGrath, Strobel, Moos,
  Retherford, and Wolven}}]{feldmanb}
Feldman, P.~D., M.~A. McGrath, D.~F. Strobel, H.~W. Moos, K.~D. Retherford, and
  B.~C. Wolven, Hst/stis ultraviolet imaging of polar aurora on ganymede,
  \textit{J. Geophys. Res.}, \textit{535}, 1085--1090, 2000.

\bibitem[{\textit{Galeev et~al.}(1986)\textit{Galeev, Kuznetsova, and
  Zeleny}}]{galeeva}
Galeev, A.~A., M.~M. Kuznetsova, and L.~M. Zeleny, Magnetopause stability
  threshold for patchy reconnection, \textit{Space Sci. Rev.}, \textit{44},
  1--41, 1986.

\bibitem[{\textit{Glocer et~al.}(2009)\textit{Glocer, T\'{o}th, Ma, Gombosi,
  Zhang, and Kistler}}]{glocera}
Glocer, A., G.~T\'{o}th, Y.~Ma, T.~Gombosi, J.-C. Zhang, and L.~M. Kistler,
  Multifluid black-adaptive-tree solar wind roe-type upwind scheme:
  Magnetospheric composition and dynamics during geomagnetic storms -- initial
  results, \textit{J. Geophys. Res.}, \textit{114}(A12203),
  doi:10.1029/2009JA014,418, 2009.

\bibitem[{\textit{Hazeltine and Meiss}(1992)}]{hazeltinea}
Hazeltine, R.~D., and J.~D. Meiss, \textit{Plasma Confinement}, Addison-Wesley
  Publishing Company, Redwood City, CA, 1992.

\bibitem[{\textit{Hesse et~al.}(1999)\textit{Hesse, Schindler, Birn, and
  Kuznetsova}}]{hessea}
Hesse, M., K.~Schindler, J.~Birn, and M.~Kuznetsova, The diffusion region in
  collisionless magnetic reconnection, \textit{Phys. Plasmas}, \textit{6},
  1781--1795, 1999.

\bibitem[{\textit{Hesse et~al.}(2001)\textit{Hesse, Birn, and
  Kuznetsova}}]{hesseb}
Hesse, M., J.~Birn, and M.~Kuznetsova, Collisionless magnetic reconnection:
  Electron processes and transport modeling, \textit{J. Geophys. Res.},
  \textit{106}, 3721--3735, 2001.

\bibitem[{\textit{Jia et~al.}(2008)\textit{Jia, Walker, Kivelson, Khurana, and
  Linker}}]{jiaa}
Jia, X., R.~J. Walker, M.~G. Kivelson, K.~K. Khurana, and J.~A. Linker,
  Three-dimensional mhd simulations of ganuymede's magnetosphere, \textit{J.
  Geophys. Res.}, \textit{113}(A06212), doi:10.1029/2007JA012,748, 2008.

\bibitem[{\textit{Jia et~al.}(2009)\textit{Jia, Walker, Kivelson, Khurana, and
  Linker}}]{jiab}
Jia, X., R.~J. Walker, M.~G. Kivelson, K.~K. Khurana, and J.~A. Linker,
  Properties of ganymede's magnetosphere inferred from improved
  three-dimensional mhd simulations, \textit{J. Geophys. Res.},
  \textit{114}(A09209), doi:10.1029/2009JA014,375, 2009.

\bibitem[{\textit{Jia et~al.}(2010)\textit{Jia, Walker, Kivelson, Khurana, and
  Linker}}]{jiac}
Jia, X., R.~J. Walker, M.~G. Kivelson, K.~K. Khurana, and J.~A. Linker,
  Dynamics of ganymede's magnetopause: Intermittent reconnection under steady
  external conditions, \textit{J. Geophys. Res.}, \textit{115}, A12,202, 2010.

\bibitem[{\textit{Karimabadi et~al.}(2011)\textit{Karimabadi, Dorelli,
  Roytershteyn, Daughton, and Chac\'{o}n}}]{karimabadic}
Karimabadi, H., J.~Dorelli, V.~Roytershteyn, W.~Daughton, and L.~Chac\'{o}n,
  Flux pile-up in collisionless magnetic reconnection: Bursty interaction of
  large flux ropes, \textit{Phys. Rev. Lett.}, \textit{107}, 025,002, 2011.

\bibitem[{\textit{Kivelson et~al.}(1998)\textit{Kivelson, Warnecke, Bennett,
  Joy, Khurana, Linker, Russell, Walker, and Polanskey}}]{kivelsonb}
Kivelson, M.~G., J.~Warnecke, L.~Bennett, S.~Joy, K.~K. Khurana, J.~Linker,
  C.~T. Russell, R.~J. Walker, and C.~Polanskey, Ganymede's magnetosphere:
  Magnetometer overview, \textit{J. Geophys. Res.}, \textit{103}, 19,963 --
  19,972, 1998.

\bibitem[{\textit{Kivelson et~al.}(2004)\textit{Kivelson, Bagenal, Kurth,
  Neubauer, Paranicas, and Sauer}}]{kivelsona}
Kivelson, M.~G., F.~Bagenal, W.~S. Kurth, F.~M. Neubauer, C.~Paranicas, and
  J.~Sauer, \textit{Jupiter: The Planet, Satellites and Magnetosphere}, chap.
  Magnetospheric interactions with satellites, pp. 513--537, Cambridge U.
  Press, 2004.

\bibitem[{\textit{Knoll and Chacon}(2006)}]{knolla}
Knoll, D., and L.~Chacon, Coalescence of magnetic islands in the
  low-resistivity, hall-mhd regime, \textit{Phys. Rev. Lett.}, \textit{96},
  135,001, 2006.

\bibitem[{\textit{Ma and Bhattacharjee}(1996)}]{mac}
Ma, Z.~W., and A.~Bhattacharjee, Fast impulsive reconnection and current sheet
  intensification due to electron pressure gradients in semi-collisional
  plasmas, \textit{Geophys. Res. Lett.}, \textit{23}, 1673--1676, 1996.

\bibitem[{\textit{Mandt et~al.}(1994)\textit{Mandt, Denton, and
  Drake}}]{mandta}
Mandt, M.~E., R.~E. Denton, and J.~F. Drake, Transition to whistler mediated
  magnetic reconnection, \textit{Geophys. Res. Lett.}, \textit{21}, 73--76,
  1994.

\bibitem[{\textit{McGrath et~al.}(2013)\textit{McGrath, Jia, Retherford,
  Feldman, Strobel, and Sauer}}]{mcgratha}
McGrath, M., X.~Jia, K.~Retherford, P.~D. Feldman, D.~F. Strobel, and J.~Sauer,
  Aurora on ganymede, \textit{J. Geophys. Res.},
  \textit{118}(doi:10.1002/jgra.50122), 2043--2054, 2013.

\bibitem[{\textit{M\"{u}ller et~al.}(2012)\textit{M\"{u}ller, Simon, Whang,
  Motschmann, Heyner, Sch\"{u}le, Ip, Keindienst, and Pringle}}]{mullera}
M\"{u}ller, J., S.~Simon, Y.-C. Whang, U.~Motschmann, D.~Heyner, J.~Sch\"{u}le,
  W.-H. Ip, G.~Keindienst, and G.~Pringle, Origin of mercury's double
  magnetopause: 3d hybrid simulation study with a.i.k.e.f., \textit{Icarus},
  \textit{218}, 666--687, 2012.

\bibitem[{\textit{Otto}(2001)}]{ottoa}
Otto, A., Geospace environment modeling (gem) magnetic reconnection challenge:
  Mhd and hall mhd -- constant and current dependent resistivity models,
  \textit{J. Geophys. Res.}, \textit{106}, 3751--3757, 2001.

\bibitem[{\textit{Parker}(1957)}]{parkerd}
Parker, E.~N., Sweet's mechanism for merging magnetic fields in conducting
  fluids, \textit{J. Geophys. Res.}, \textit{62}, 509, 1957.

\bibitem[{\textit{Paty and Winglee}(2004)}]{patya}
Paty, C., and R.~Winglee, Multi-fluid simulations of ganymede's magnetosphere,
  \textit{Geophys. Res. Lett.}, \textit{31}(L24806), doi:10.1029/2004GL021,220,
  2004.

\bibitem[{\textit{Paty and Winglee}(2006)}]{patyb}
Paty, C., and R.~Winglee, The role of ion cyclotron motion at ganymede:
  Magnetic field morphology and magnetospheric dynamics, \textit{Geophys. Res.
  Lett.}, \textit{33}(L10106), doi:10.1029/2005GL025,273, 2006.

\bibitem[{\textit{Paty et~al.}(2008)\textit{Paty, Paterson, and
  Winglee}}]{patyc}
Paty, C., W.~Paterson, and R.~Winglee, Ion energization in ganymede's
  magnetosphere: Using multifluid simulations to interpret ion energy
  spectrograms, \textit{J. Geophys. Res.}, \textit{113}(A06211),
  doi:10.1029/2007JA012,848, 2008.

\bibitem[{\textit{Petschek}(1964)}]{petscheka}
Petschek, H.~E., Magnetic field annihilation, {\it AAS-NASA Symposium on
  Physics of Solar Flares, NASA Spec. Publ. 50}, 425 (1964), 425-439, 1964.

\bibitem[{\textit{Powell et~al.}(1999)\textit{Powell, Roe, Linde, Gombosi, and
  Zeeuw}}]{powella}
Powell, K.~G., P.~L. Roe, T.~J. Linde, T.~I. Gombosi, and D.~L.~D. Zeeuw, A
  solution-adaptive upwind scheme for ideal magnetohydrodynamics, \textit{J.
  Comp. Phys.}, \textit{154}, 284--309, 1999.

\bibitem[{\textit{Rogers et~al.}(2001)\textit{Rogers, Denton, Drake, and
  Shay}}]{rogersa}
Rogers, B.~N., R.~E. Denton, J.~F. Drake, and M.~A. Shay, The role of
  dispersive waves in collisionless magnetic reconnection, \textit{Phys. Rev.
  Lett.}, \textit{87}, 195,004, 2001.

\bibitem[{\textit{Shay and Drake}(1998)}]{shayb}
Shay, M.~A., and J.~F. Drake, The role of electron dissipation on the rate of
  collisionless magnetic reconnection, \textit{Geophys. Res. Lett.},
  \textit{25}, 3759--3762, 1998.

\bibitem[{\textit{Shay et~al.}(1999)\textit{Shay, Drake, and Rogers}}]{shaya}
Shay, M.~A., J.~F. Drake, and B.~N. Rogers, The scaling of collisionless,
  magnetic reconnection for large systems, \textit{Geophys. Res. Lett.},
  \textit{26}, 2163--2166, 1999.

\bibitem[{\textit{Slavin et~al.}(2008)}]{slavina}
Slavin, J.~A., et~al., Mercury's magnetosphere after messenger's first flyby,
  \textit{Science}, \textit{321}, doi:10.1126/science.1159,040, 2008.

\bibitem[{\textit{Sweet}(1958)}]{sweeta}
Sweet, P.~A., The neutral point theory of solar flares, in
  \textit{Electromagnetic Phenomena in Cosmic Physics}, edited by B.~Lehnert,
  pp. 123--134, Cambridge University Press, London, 1958.

\bibitem[{\textit{Tanaka}(1994)}]{tanakaa}
Tanaka, T., Finite volume tvd scheme on an unstructured grid system for
  three-dimensional mhd simulations of inhomogeneous systems including strong
  background potential field, \textit{J. Comput. Phys.}, \textit{111}(381),
  1994.

\bibitem[{\textit{T\'oth et~al.}(2008)\textit{T\'oth, Ma, and Gombosi}}]{totha}
T\'oth, G., Y.~Ma, and T.~I. Gombosi, Hall magnetohydodynamics on
  block-adaptive grids, \textit{J. Comp. Phys.}, \textit{227}, 6967--6984,
  2008.

\bibitem[{\textit{Winglee}(1994)}]{wingleea}
Winglee, R.~M., Non-mhd influences on the magnetospheric current system,
  \textit{J. Geophys. Res.}, \textit{99}(A7), 13,437--13,454, 1994.

\bibitem[{\textit{Winglee}(2004)}]{wingleeb}
Winglee, R.~M., Ion cyclotron and heavy ion effects on reconnection in a global
  magnetotail, \textit{J. Geophys. Res.}, \textit{109}(A09206),
  doi:10.1029/2004JA010,385, 2004.

\end{thebibliography}

\end{article}

\end{document}